\documentclass[aps,pre,preprint]{revtex4-1}%
\usepackage{amsfonts}
\usepackage{amsmath}
\usepackage{amssymb}
\usepackage{graphicx}
\usepackage{natbib}
\usepackage[caption=false]{subfig}
\providecommand{\U}[1]{\protect\rule{.1in}{.1in}}

\DeclareMathOperator{\sgn}{sgn}

\begin{document}
\preprint{ }
\title{Hydration of ions in 2D water}

\author{S. Dutta}
\affiliation{ Asia Pacific Center for Theoretical Physics, Pohang, Gyeongbuk, 790-784, Korea}
\author{Yongjin Lee and Y.S. Jho}
\email{ysjho@apctp.org}
\affiliation{Department of Physics, Pohang University of Science and Technology, Asia Pacific Center for Theoretical Physics, Pohang, Gyeongbuk, 790-784, Korea}

\begin{abstract}
We present a 2D lattice model of water to study the effects of ion hydration on the properties of water. 
We map the water molecules as lattice particles consisting of a single Oxygen at the center of a site and two Hydrogen atoms 
on each side. 
The internal state of the system, such as the dipole moment at a site, is defined with respect to the location of the Hydrogen atoms at the site
depending on their role in Hydrogen bonds (H-bonds) being a donor or an acceptor. We study the influence of the charge and the radius of the ion 
on the insertion energy and on the H-bonds in the first and second hydration layers around the ion and in the bulk. 
In particular we analyze how the competing interactions of the short-ranged H-bonds and the long-ranged electrostatics influence 
the hydration properties. The role of the ion both as a source of the electrostatic interactions as well as a defect is also discussed. 
Our model also shows the well known fact that the polarizability of the water molecules destroys the hydrogen bond network and 
increases the dipole moment of the molecules near the ion.
\end{abstract}


\maketitle

\section{Introduction}
Interactions of ions with water form an important component of all biological and chemical systems~\cite{chaplin2006we}. Ions change the solubility of proteins and nucleic acids in water. This in turn 
changes their folding, permeation and self-assembly properties~\cite{israelachvili2011intermolecular}. In chemical systems they change the rate of reactions, and in physics,
they change not only the static properties but also the dynamic properties such as diffusion and transport coefficients. This in turn changes the rheology and  hydrodynamics
of water molecules \cite{chaplin2009theory}. Hydration of ions is
a vast field that has been studied extensively through experiments and theories \cite{petersen2006nature,marcus2009effect}.

Ions not only modify the structure and the orientations of the water molecules through the long-ranged electrostatic forces, they help in making or breaking of the short-ranged 
H-bonding, especially the first hydration shell \cite{omta2003negligible,salis2014models,marcus2009effect}. Introducing an ion in the solution changes the existing structure of the H-bond network as the water molecules reorient themselves and break the H-bonds with 
mostly their nearest neighbors, and sometimes with their second nearest neighbors. Beyond the hydration layer of the ions, the electric field is screened by the dielectric constant, $\epsilon$. Several experiments such as infrared and 
Raman spectroscopic measurements, neutron diffraction spectroscopy and X-ray absorption spectroscopy have been performed to investigate the effect of the ions on the 
H-bond network \cite{ohtaki1993structure}. Simulations of water may be more tractable to observe the detail of the H-bond structure \cite{jungwirth2006specific}.
But the simulation results depend strongly on the water model used. Many of the results from these experiments and simulations are contradict each other. Theoretical 
modeling of water can capture the underlying physics of the ion hydration and is essential to understand their effects on the water structure.
Lynden-Bell and Rasaiah \cite{lynden1997hydrophobic} developed a method to determine thermodynamic properties of hydrated solutions by treating the charge
and the size of the ions as dynamical variables. Using this method they could move smoothly from hydrophilic to hydrophobic solvation conditions.
Classical Density Functional Theory (DFT) and integral equation theories have also been employed in the literature to study water properties \cite{prendergast2005electronic,urbivc2002two}. 
However most of these theories can not predict the details of the H-bonding structure due to the complexity of the orientation dependent interactions. 
Collins \cite{Collins} suggested that hydration effects on water can be described by a competition between the ion-water electrostatic interactions and the water-water
H-bonding interactions. Hribar \textit{et al.} \cite{Hribar} used 2D MB water model \cite{dill2005modeling} to model ion solvation in water. They take into account the electrostatics of the 
ion solution by treating the water molecules as dipoles and the H-bonds as short-ranged Gaussian bonds. Their model predicts that the H-bonds break more easily for smaller ions than larger ones. 

Indeed, to understand the ion specific effect, we have to consider the complexity of the ionic water, in which the rather long-ranged Coulomb interaction is in a
subtle balance with the short-ranged H-bond attraction. The structural change due to the introduction of the ion should be fully considered in order to reflect this subtle change. 
The simplest way to consider the structure is to construct a lattice model of water~\cite{hone2006lattice}. In this work, we develop a lattice based water model in 
two dimensions, with the water molecules located at the sites of a square lattice. Each Oxygen at a site has four sub-sites, of which two are occupied by Hydrogen atoms
as shown in Figure \ref{Fig1.1}.
We disregard the case when the Hydrogen atoms are situated opposite to each other. Thus each water molecule has a $90^{\circ}$ structure. 
Since the water molecules are fixed at a given position our model has two kind of interactions - the long ranged electrostatic ion-water interactions and
the orientational correlations caused by the H-bonding. The location of the Hydrogen atoms at a water site with respect to the Oxygen is described 
using the Ising states. The short ranged H-bonds are then described by the nearest neighbor interactions and the ion-water interactions by the monopole-dipole electrostatic interactions. 
Lattice models in 1D have been successfully applied to water problems before in Refs \cite{nanowaterJCP2009,nanowaterNJP2010,nanowaterPRL2009,nanowaterPNAS2008}
to describe water defects in nanowires. However in 1D the angular structure of water is missing, and the competing effect between different interactions disappears, 
and hence it is not suitable for hydration shell study. 2D is the minimal dimension that contains these effects. 
In Section \ref{Sec1}, using the 2D Ising Hamiltonian, we calculate 
the number of H-bonds, which is a good measure of network structure~\cite{marcus2009effect}. The insertion energy and the average orientation order in the first and second hydration shell and in the bulk are also calculated. 
In Section \ref{Sec2} we perform mean field calculations of our model with a effective potential and compare our results with the existing experimental and simulation results. 
We study the differences between the hydration of anions and cations in Section \ref{Sec3} by explicitly considering the Hydrogen and Oxygen atoms as point charges instead of dipoles.
In Section \ref{Sec4} we study the behavior of the H-bonds in the first and second hydration shell with the variation in the charge and the radius of the ions. 
Finally in Section \ref{Sec5}
the polarizability of the water molecules is considered and its effects on the H-bonds and the dipole moment of the molecules are discussed. 

\section{Lattice model}
\label{Sec1}
\begin{figure*}
    \includegraphics[trim={5cm 3cm 5cm 3cm},clip,scale=0.6]{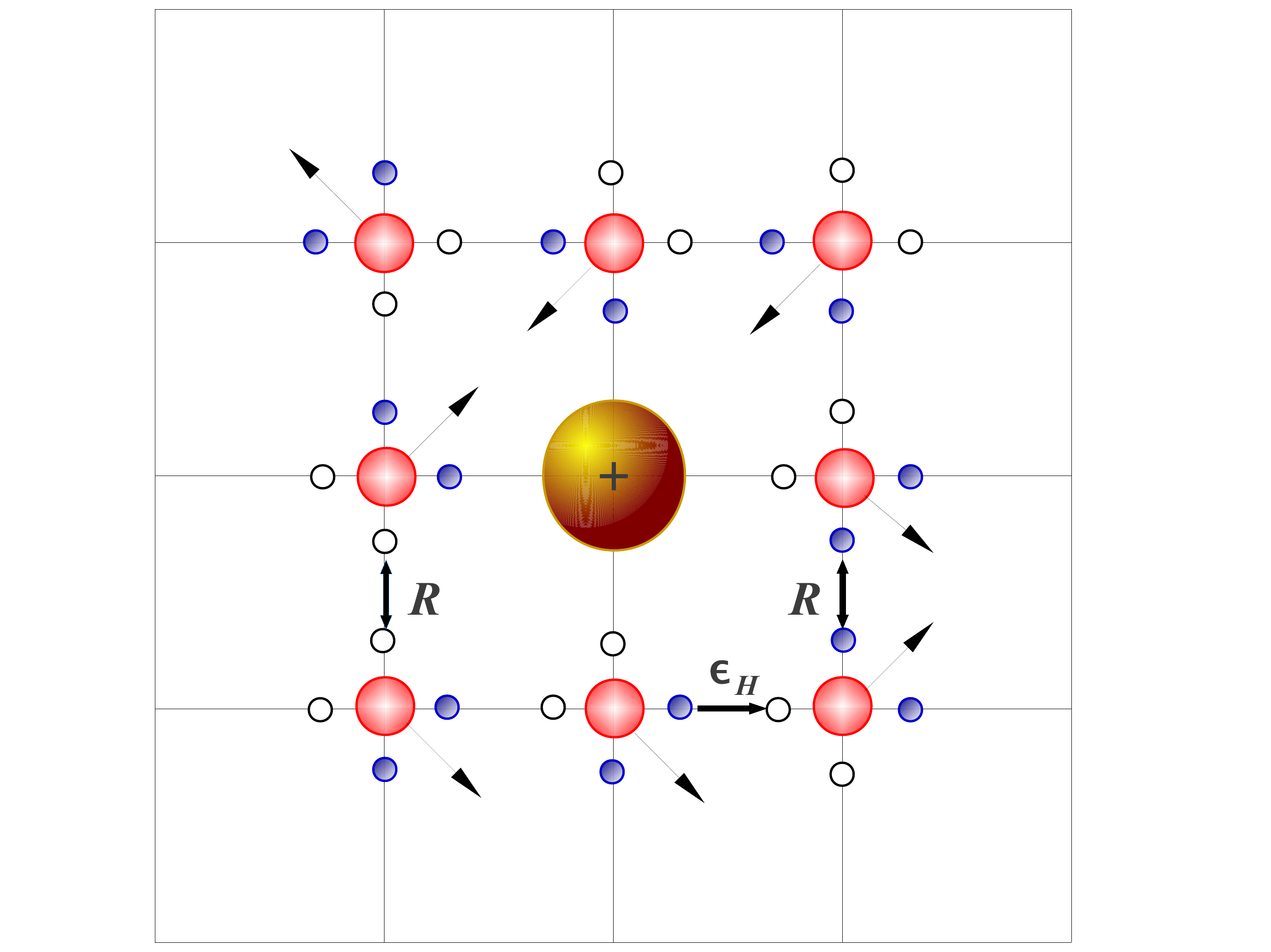}
    \caption{(Color online) 2D lattice model for water with an ion at the center. Each lattice site has one Oxygen atom at the center and four sites, two of which are occupied 
    by Hydrogen atoms (blue dots) and the other two are 
    unoccupied. Also shown are the two interactions, the repulsive $R$ interactions and the attractive H-bonding interactions $\epsilon_H$. }
    \label{Fig1.1}
\end{figure*}

We consider $N^2-1$ water molecules arranged on a 2D square lattice, with $N$ rows and $N$ columns, as shown in Figure \ref{Fig1.1}. Each lattice site has an Oxygen atom at the center 
and two Hydrogen atoms. We assume the Hydrogen atoms occupy the consecutive sites only at each lattice points so that we have non-zero dipole moment for each water molecule. 
Thus at each site there is one Hydrogen along the row and one along the column denoted by superscripts $x$ and $y$ respectively. Each water molecule acts like a dipole given by the vector
$\frac{\mu_0}{\sqrt{2}}(\sigma_{i,j}^x, \sigma_{i, j}^y)$, where $\mu_0$ is the permanent dipole moment of water. An ion is located at the center of the lattice at $(N/2, N/2)$. 
The Hydrogen occupancy is defined in terms of the Ising-like states 
\begin{align}
 \sigma_{i,j}^{x,y} & = +1 & \text{H atom left or top of Oxygen atom at site $(i, j)$}, \nonumber\\
 & = -1 & \text{H atom located right or bottom of Oxygen atom at site $(i, j)$}.
 \label{eq1.1}
\end{align}
Each water molecule interacts only with its nearest neighbor. Since we consider finite number of water molecules we use the periodic boundary conditions to mimic bulk water. 
The interactions are repulsive $R$ if the two nearest sites are occupied by Hydrogen atoms and attractive (Hydrogen bond) $\epsilon_H$ if one of the
two nearest sites are unoccupied as shown in the Figure \ref{Fig1.1}. The Hydrogen bond attractions are about 3 to 4 times stronger than the repulsions between the
Hydrogen sites. The typical Hydrogen bond strength is known about $-10k_BT \sim -5 k_BT$. However, this is too strong for our model as the system freezes at those conditions. 
Instead, we choose smaller value for the Hydrogen bond strength $\epsilon_H$ as $-4k_BT$ and $-2k_BT$ and repulsions $R$ as $k_BT$ and $0.5k_BT$ respectively. 
We introduce an ion to the system. At each water site we get an additional energy contribution given by a dipole-ion interaction 
$-\boldsymbol{\mu}_{ij}.\mathbf{E}_{ij} = -\frac{\Gamma}{(r(i,j)-R_0)^2}\sigma_{ij}$, where $r(i,j)$ is the position of the ion at the water site $(i, j)$ and $R_0$ is the position of the ion.
In the present calculations we consider $\Gamma = 1k_BT$ and $10k_BT$ respectively. The ion-dipole interactions are long-ranged $r^{-2}$, hence we need to sum over the periodic images of the ion with the periodicity $Na$. 

The dimensionless Hamiltonian of the system is 
\begin{align}
 \beta H_N & = \frac{\epsilon_H}{2}\sum_{i, j}(\sigma_{i+1, j}^x\sigma_{i, j}^x+1) + \frac{R}{2}\sum_{i, j}(\sigma_{i+1, j}^x\sigma_{i, j}^x-1) 
 + \frac{\epsilon_H}{2}\sum_{i, j}(\sigma_{i, j+1}^y\sigma_{i, j}^y+1) \nonumber\\ & + \frac{R}{2}\sum_{i, j}(\sigma_{i, j+1}^y\sigma_{i, j}^y-1) 
  - \sum_{i, j}^{\prime}\frac{\Gamma}{(\vert i-N/2\vert^2 + \vert j-N/2\vert^2)^{3/2}}\left(\vert i-N/2\vert\sigma_{i,j}^x+\vert j-N/2\vert\sigma_{i,j}^y\right)\nonumber\\
 & = J/2\sum_{i,j}\left(\sigma_{i+1,j}^x\sigma_{i, j}^x + \sigma_{i,j+1}^y\sigma_{i, j}^y\right) - \sum_{i, j}\Gamma \left(g_x(i, j)\sigma_{i, j}^x + g_y(i, j)\sigma_{i, j}^y\right)
 + N^2(\epsilon_H-R), 
 \label{eq1.3}
\end{align}
where $J = \epsilon_H + R$ is the strength of the interactions. The $\prime$ in the summation in the last term implies the summation over the image charges of the ion.
The ion-water site interactions are $ g_x(i, j) = \vert i-N/2\vert/(\vert i-N/2\vert^2 + \vert j-N/2\vert^2)^{3/2}$
and $ g_y(i, j) = \vert j-N/2\vert/(\vert i-N/2\vert^2 + \vert j-N/2\vert^2)^{3/2}$. 

\section{Ion hydration shell}
\label{Sec2}
\begin{figure*}[h]
        \centering
           \subfloat{%
              \includegraphics[scale=0.5]{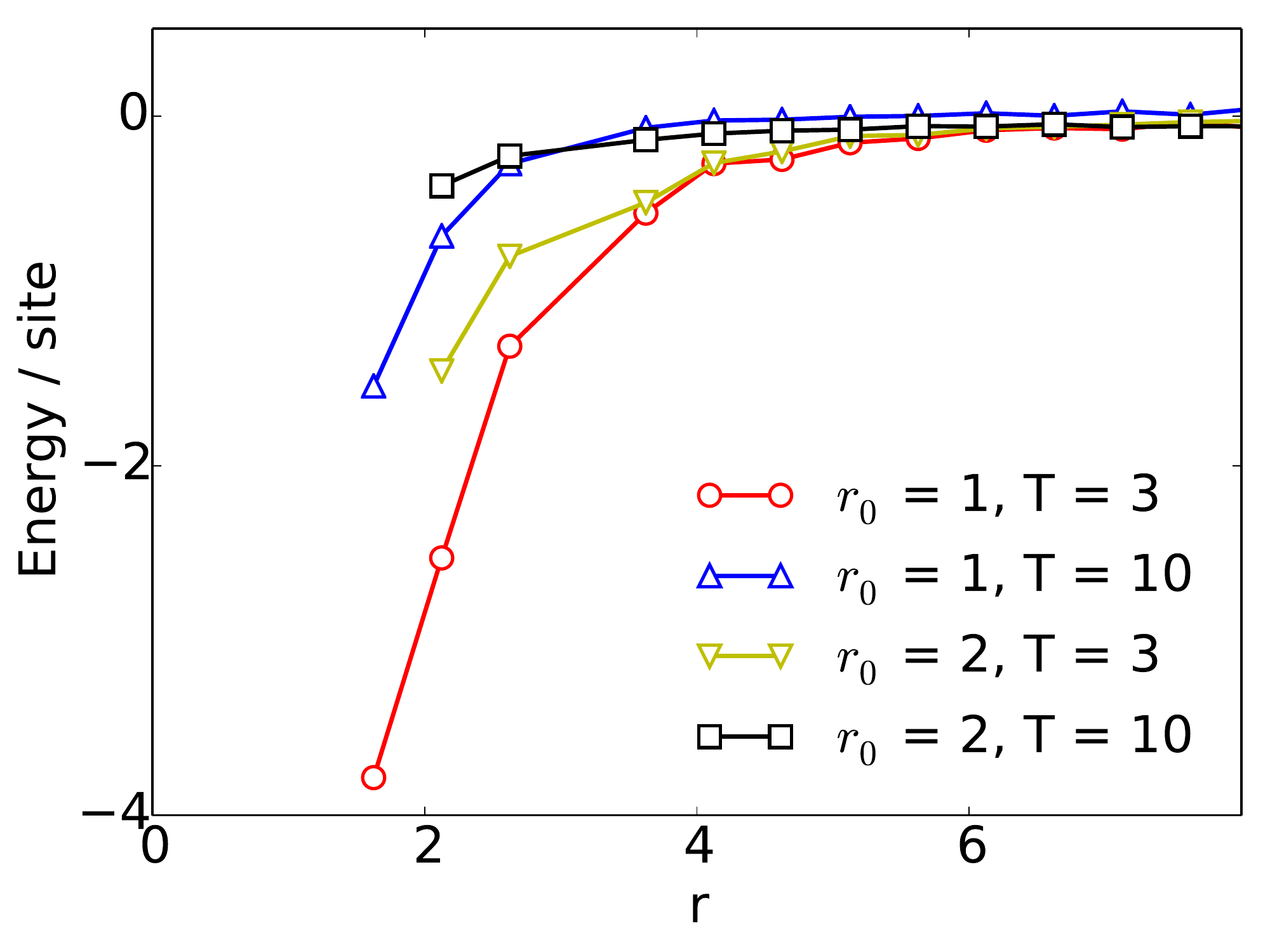}%
              \label{Fig2.1b}%
           }
           \caption{(Color online) Insertion energy at each site  vs distance of the water site from the ion for the strength of
           the ion-dipole interactions $\Gamma = 10$.}
           \label{Fig2.1}
 \end{figure*}

\begin{figure*}[h]
        \centering
           \subfloat{%
              \includegraphics[height=6.2cm]{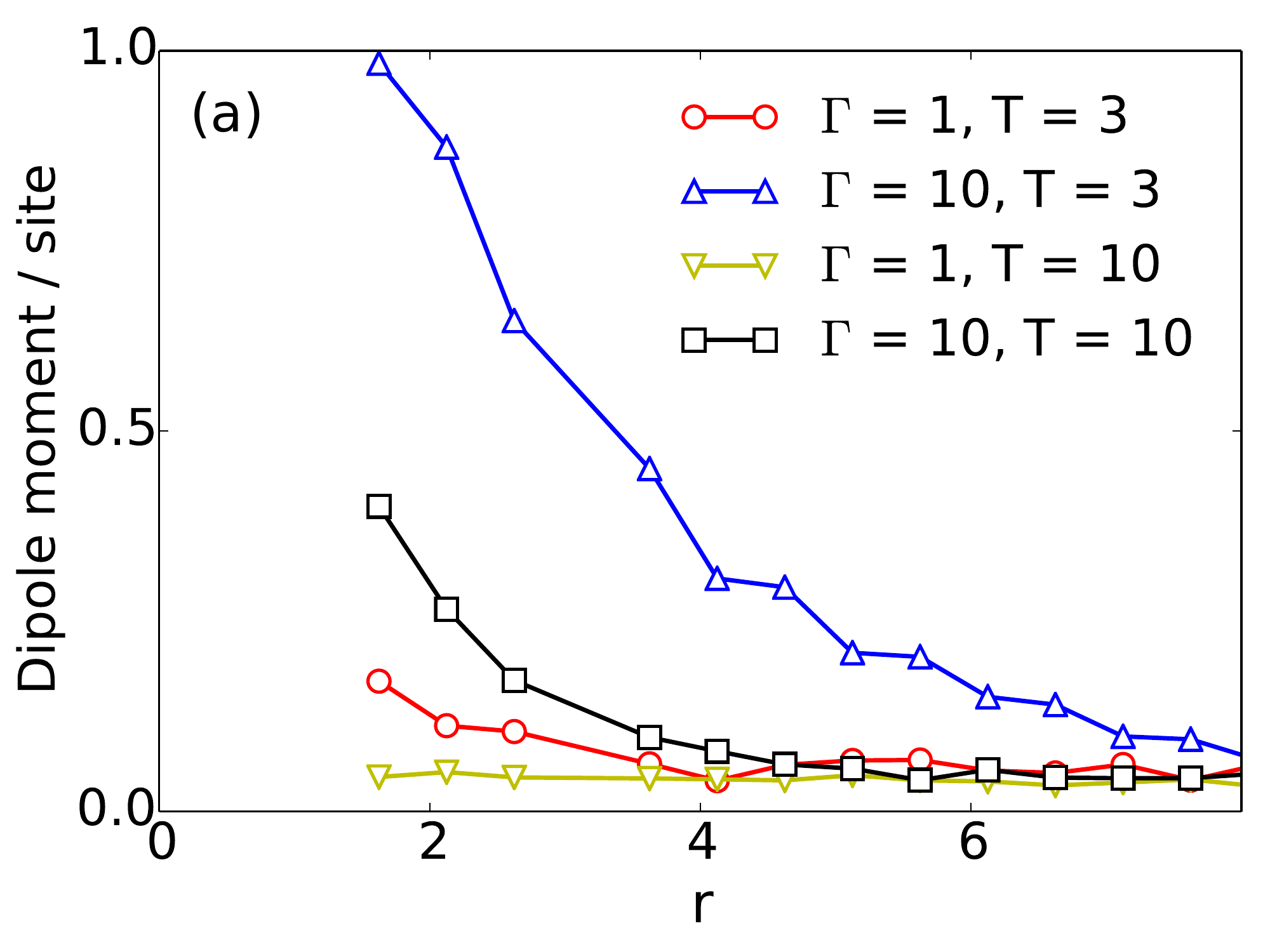}%
              \label{Fig2.3a}%
           }
           \subfloat{%
              \includegraphics[height=6.2cm]{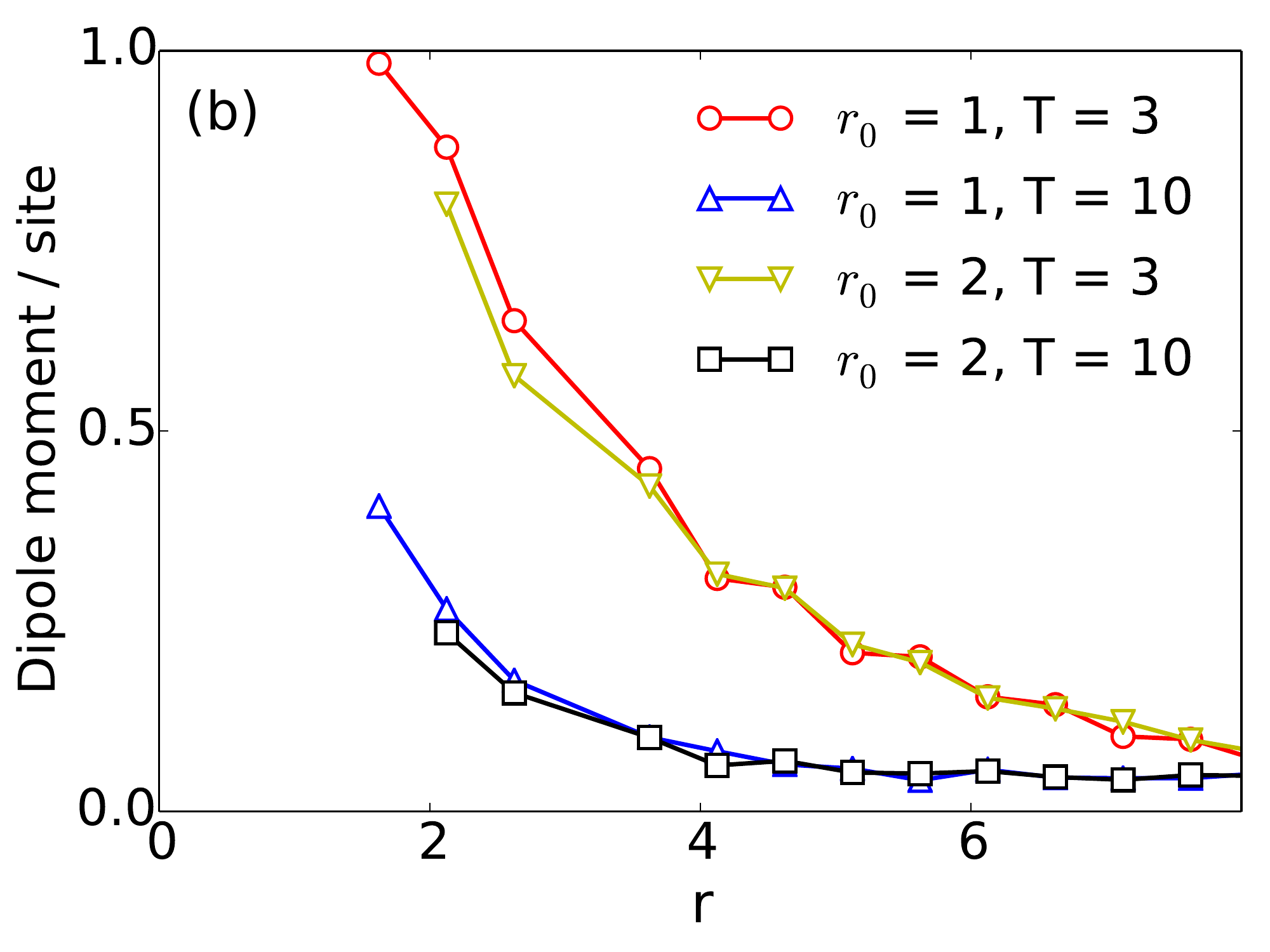}%
              \label{Fig2.3b}%
           }
           \caption{(Color online) Dipole at each site vs distance of the site from the ion at fixed (a) ion radius, $r_0 = 1$ and (b) $\Gamma = 10$.}
           \label{Fig2.3}
 \end{figure*}

We define an insertion energy as the energy difference to add an ion to the water with respect to the bulk water energy.
More negative the insertion energy of the ion is, more likely it is for the ion to dissolve into the water. Close to the ion the insertion energy at each site depends 
strongly on the radius of the ion, $r_0$ and the strength of the  ion potential as shown in Figure \ref{Fig2.1}. At distances further from the ion, this energy goes to zero as
expected. In this work we only consider dimensionless quantities. The energies are scaled with respect to $k_BT$, the distances like the ion-water distance $r$ 
and the radius of the ion $r_0$ with respect to the lattice constant which is $3 \AA$ in our model and the temperature is scaled with respect to the room temperature $298$K.  
When the ion strength (charge) is weak the insertion energy is low and the ion is less likely to be dissolved in the water.
The cost of cavity energy exceeds the energy benefit of the ion-water interactions. 
When the ion is highly charged the electrostatic ion-water interactions overcome the cavity energy to disrupt the water-water interaction and the ion finds it 
easier to dissolve in the water. Similarly from Figure \ref{Fig2.1} we see that as the radius of the ions
increases the insertion energy becomes less negative and hence less likely to dissolve. Note that we are talking about the energy not free energy as
the positional entropy is zero. The quantities $\sigma^x$ and $\sigma^y$ defined in the previous Section \ref{Sec1} denote the alignment state of a water
dipole at a lattice site. After thermal averaging we get the average dipole moment at each site, $\frac{\mu_0}{\sqrt{2}}\left(\langle\sigma_{i,j}^x\rangle, 
\langle\sigma_{i,j}^y\rangle\right)$ which measures the orientation order of the water dipoles. Henceforth we call the thermal averaged dipole moment
simply the dipole at a given site.
When the dipoles are fully aligned it takes the values $\pm1$ and a number between $-1$ and $1$ otherwise. The dipole moment of the water in Figures \ref{Fig2.3}-(a) 
and Figure \ref{Fig2.3}-(b) show that the orientation order among the dipoles is highest close to the ion and drops to the bulk value which is zero at high temperatures. 
Higher ion charge, smaller ion radius and low temperatures favor the alignment of the dipoles as shown in Figure \ref{Fig2.3}-(a) and Figure \ref{Fig2.3}-(b).
 
\begin{figure}[h]    
\begin{minipage}[t]{0.45\textwidth}
\subfloat{
\includegraphics[height=5.85cm]{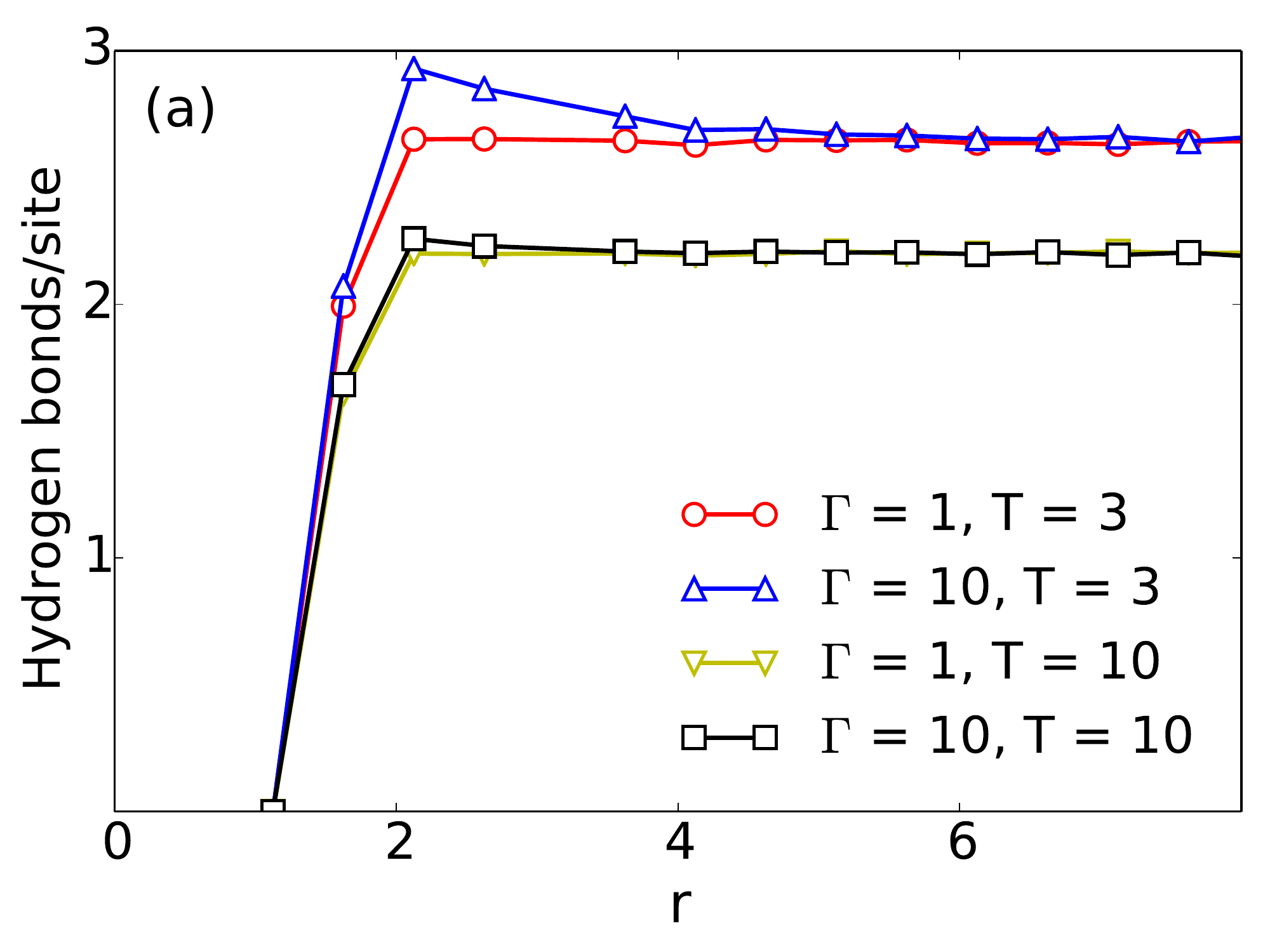}}
\end{minipage}
\hspace{5mm}
\begin{minipage}[t]{0.45\textwidth}
\subfloat{
\includegraphics[height=5.85cm]{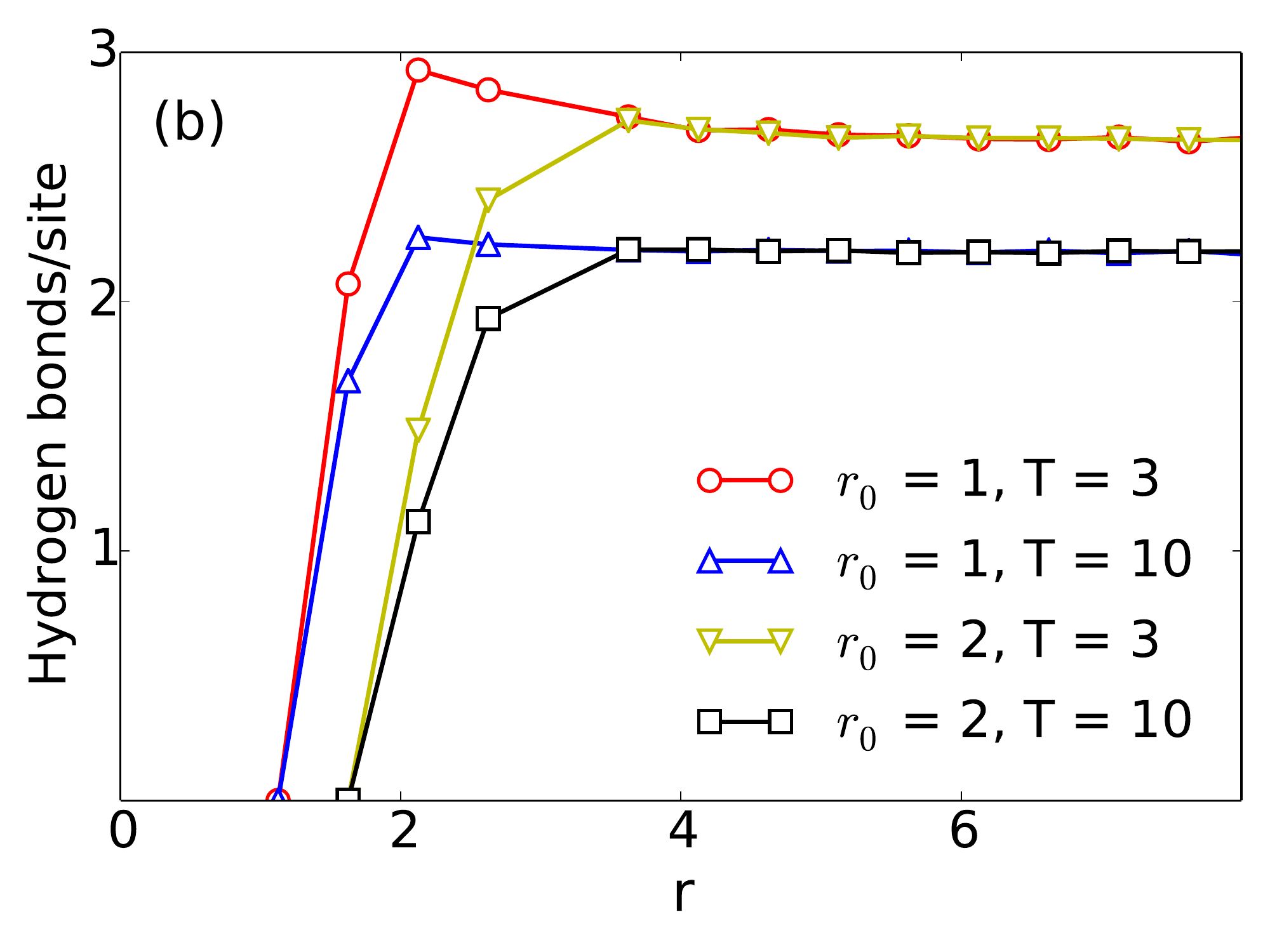}}
\end{minipage}

\vspace*{0.1cm} 
\begin{minipage}[t]{0.45\textwidth}
\subfloat{
\includegraphics[trim={0.5cm 0cm 0cm 0cm},clip,height=6cm]{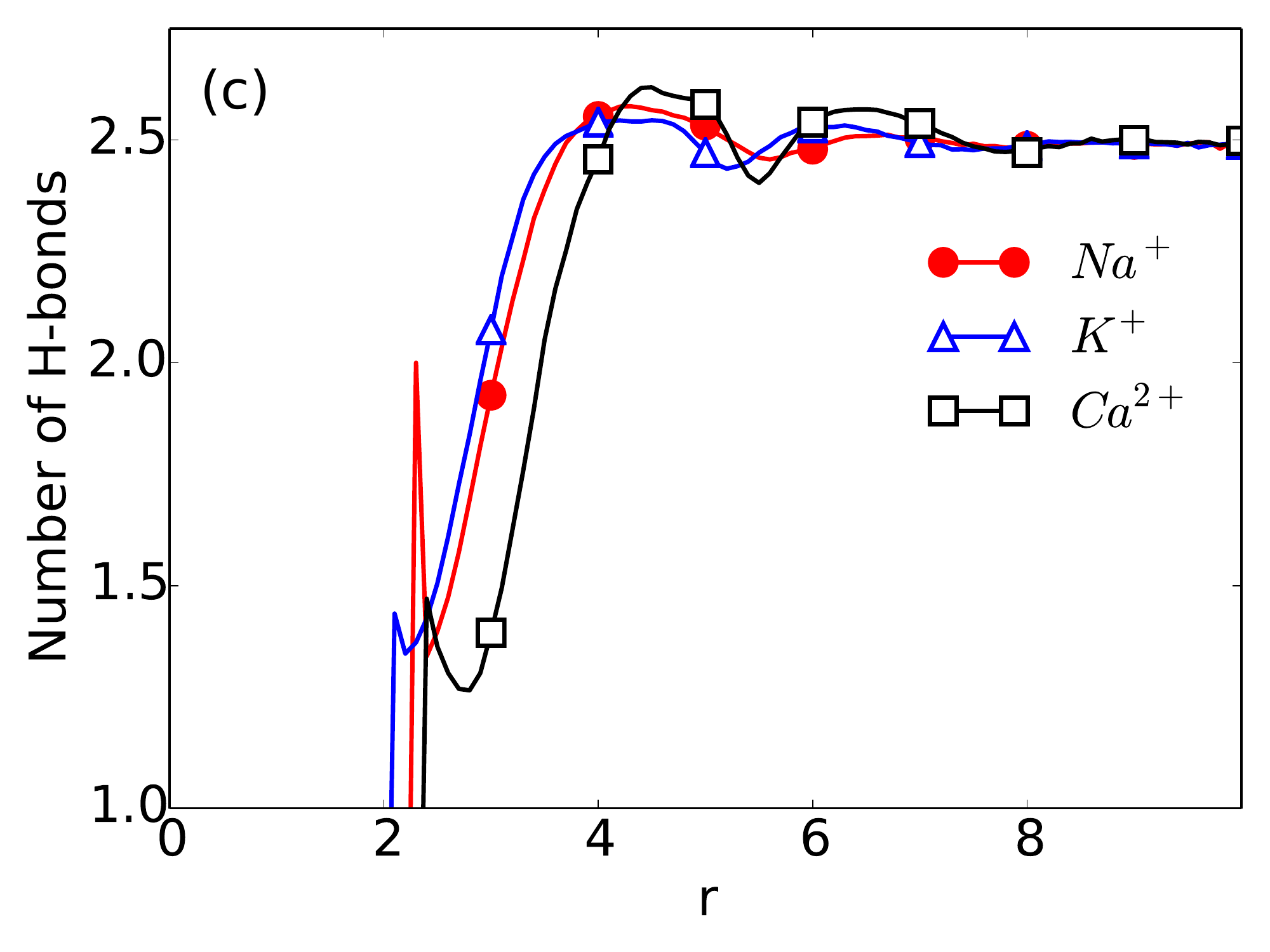}}
\end{minipage}
\hspace{4.3mm}
\begin{minipage}[t]{0.45\textwidth}
\subfloat{
\includegraphics[trim={0.5cm 0cm 0cm 0cm},clip,height=6cm]{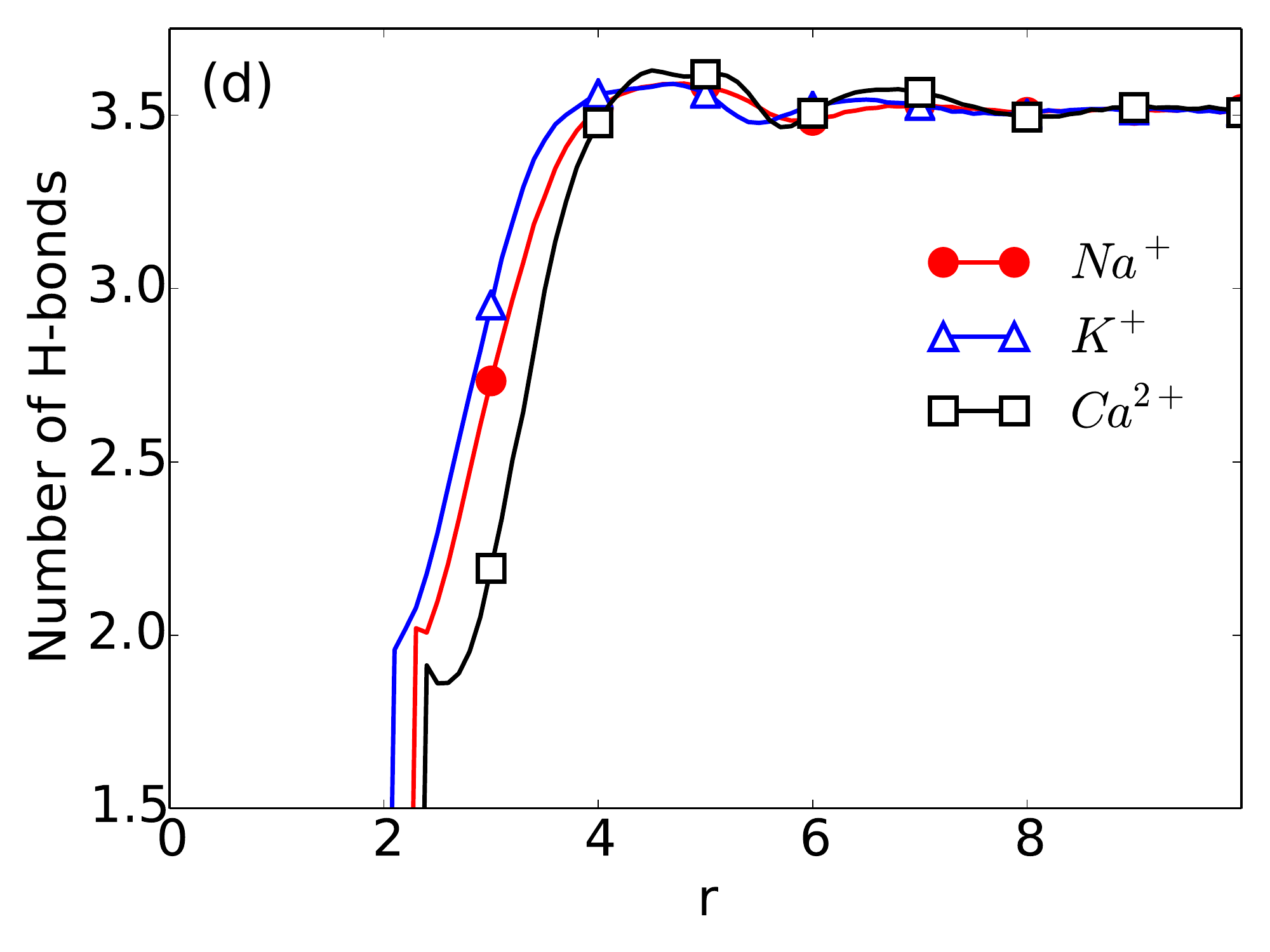}}
\end{minipage}
\caption{(Color online) H-bonds at each site vs distance of the site from the ion at fixed (a) ion radius $r_0 = 1$ and (b) $\Gamma = 10$.
           (c) H-bonds obtained Molecular Dynamics simulations for three different kind of ions, sodium $Na^{+}$, potassium $K^{+}$
           and Calcium $Ca^{2+}$ with the cutoff distance for tighter H-bonds $1.97 \AA$. (d) H-bonds with cutoff distance $3.5 \AA$.}
           \label{Fig2.2}
\end{figure}

The number of H-bonds at a given site $(i, j)$ in our model is counted by the formula
\begin{equation}
 n_H = \frac{\sigma_{i, j}^x}{2}\left(\sigma_{i-1,j}^x + \sigma_{i+1, j}^x\right) + \frac{\sigma_{i, j}^y}{2}\left(\sigma_{i,j-1}^y + \sigma_{i, j+1}^y\right) + 2.
 \label{eq2.1}
\end{equation}
It is known that the number of H-bond (the extent of H-bond) has been regarded a measure that represents the structural properties well~\cite{marcus2009effect}. 
We find from our lattice model that the nearest shell of the ion has a lower number of H-bonds than that of the bulk water. While the next immediate layer have
higher number of H-bonds than that of the bulk. This can be understood from the fact that we get most number of H-bonds when the water dipoles are aligned 
along a specific direction. This is because of our definition of H-bonds in the equation \eqref{eq2.1} as the orientational correlations.
Although the first layer has less number of H-bond due to the strong directional force from the ion and the deficit of bonding sites due to the presence of the ion, the H-bond at the next sites can be even larger because the alignment of the hydrogens in first layer acts as a biased boundary condition to second layer. Our simple model captures this effect as shown in the
Figure \ref{Fig2.2}-(a) and Figure \ref{Fig2.2}-(b). 

Because of the excluded volume of the ion there is no H-bond inside the ion radius. The second water layer has the highest number of 
H-bonds as shown by the slight bump in the plot. From Figure \ref{Fig2.2}-(a) and Figure \ref{Fig2.2}-(b) we see that larger ion charge and smaller ion size would 
produce more H-bonds. To test our theory we perform Molecular Dynamics simulations in AMBER with POL3 water model. We consider three kind of ions, Sodium,
Potassium and Calcium, with a very dilute concentration of 1M dissolved in 890 water molecules. We have performed NPT simulations at $300K$ and pressure $1.01325$ bar. 
After energy minimization,
we ran the simulations for $10.320$ ns for the system to equilibrate. We have plotted our data for the number of H-bonds vs the distance from the ions
( Sodium, Potassium or Calcium ) in Figures \ref{Fig2.2}-(c) and \ref{Fig2.2}-(d). In our simulation two water molecules are considered to be H-bonded if the shortest 
distance between an oxygen in one molecule and a hydrogen in the other molecule is less than a cutoff distance. In Figure \ref{Fig2.2}-(c), the cutoff is chosen 
to be $1.97 \AA$ and in Figure \ref{Fig2.2}-(d) $3.5 \AA$. From the simulations we find that second shell has higher number of H-bonds compared to the first shell as obtained from the theory.
$Na^{+}$ and $K^{+}$ have the same valence but ionic radius of $K^{+}$ is $1.3$ times that of $Na^{+}$. On the other hand $Ca^{2+}$ and $Na^{2+}$ have almost
the same ionic radius but valence of $Ca^{2+}$ is twice of $Na^{+}$. For this reason the ion-water interaction in first shell is weaker for $K^+$ than $Na^{+}$ which leads 
to the more H-bonds at first shell. But, the differences in their second or third shells are not significant.  
In contrast, the number of the H-bonds in the first shell is almost the same for $Na^+$ and $Ca^{2+}$ in spite of their difference in valence (their ionic radii are 
almost the same.). However there is a substantial difference in H-bonds at their second shells. Not only is the number of the H-bonds in the second shell larger for $Ca^{2+}$, 
but also the shell is sharper due to the stronger H-bond network. The stronger network for second shell is coming from the first shell configuration which 
is supposed to be more biased for $Ca^{2+}$ because of the stronger ion-water interaction at first shell (Because the ion-water interaction decreases fast as distance, 
their contribution at second shell becomes much smaller than the interaction strength of the H-bonds.). This behavior is also seen in Figure \ref{Fig2.2}-(a) where 
increasing $\Gamma$ increases the H-bonds in the second shell more significantly than the first shell. 

\begin{figure*}[h]
        \centering
           \subfloat{%
              \includegraphics[height=4.2cm]{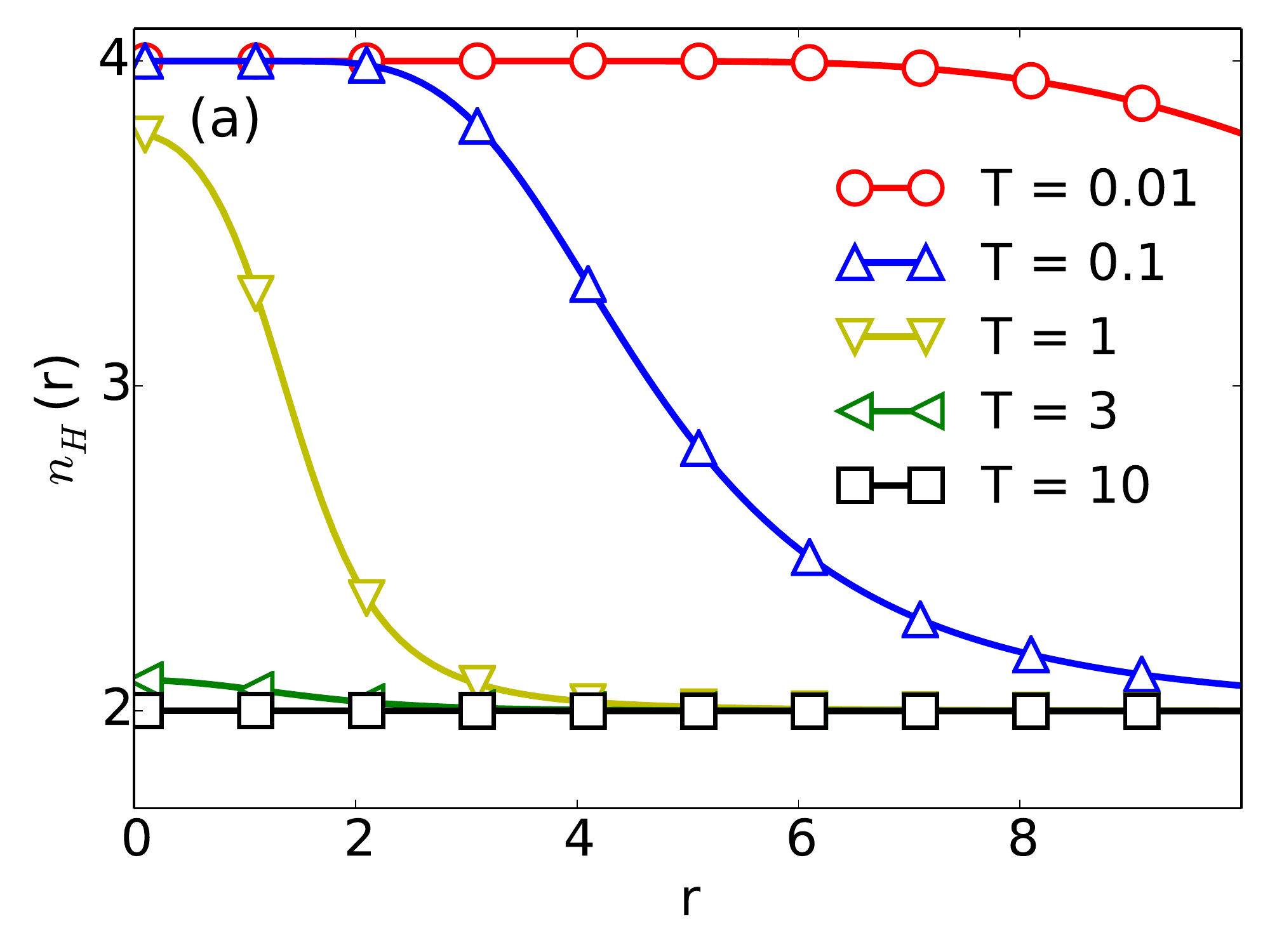}%
              \label{Fig2.4a}%
           }
           \subfloat{%
              \includegraphics[height=4.2cm]{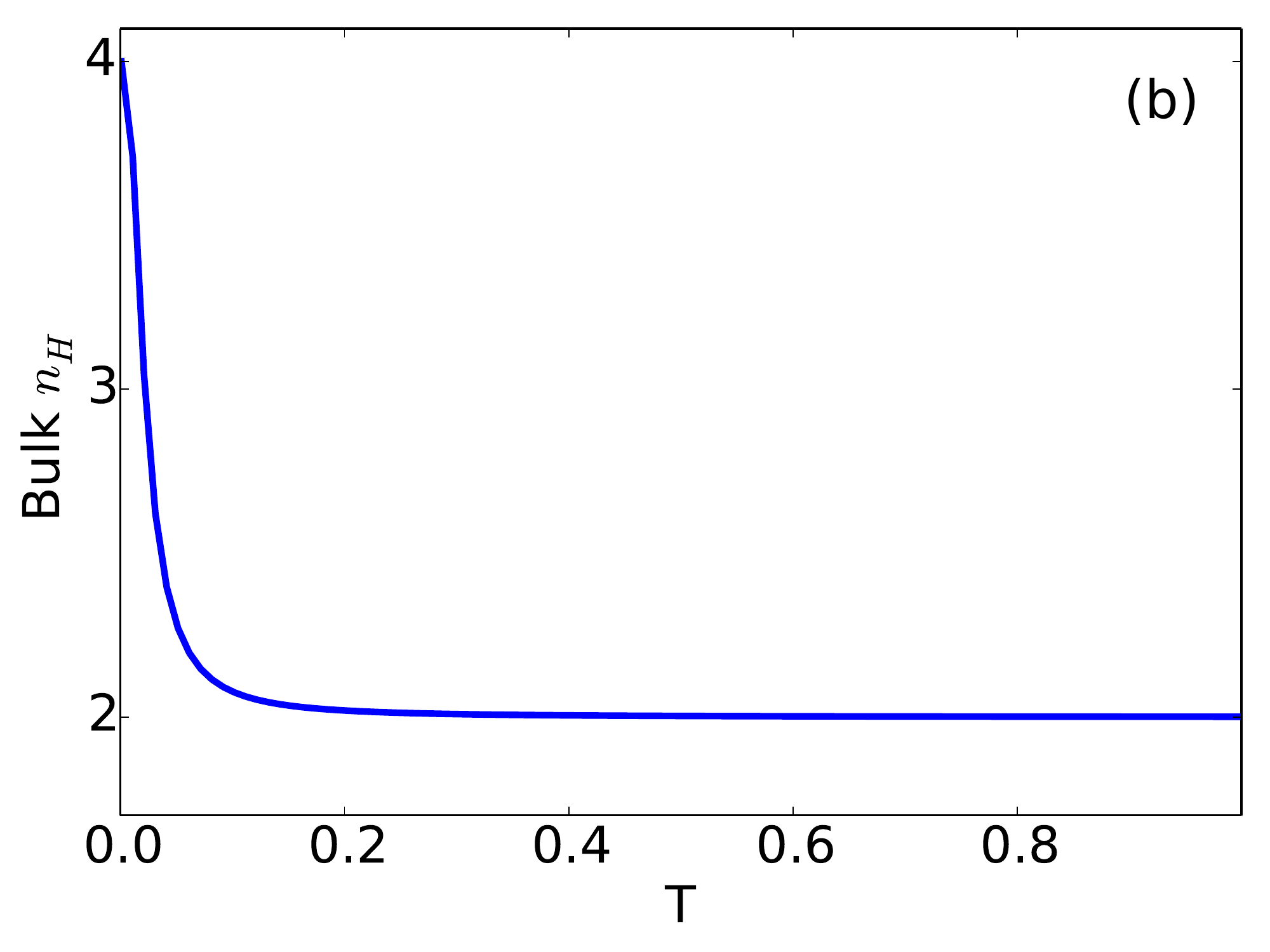}%
              \label{Fig2.4b}%
           }
           \subfloat{%
              \includegraphics[height=4.2cm]{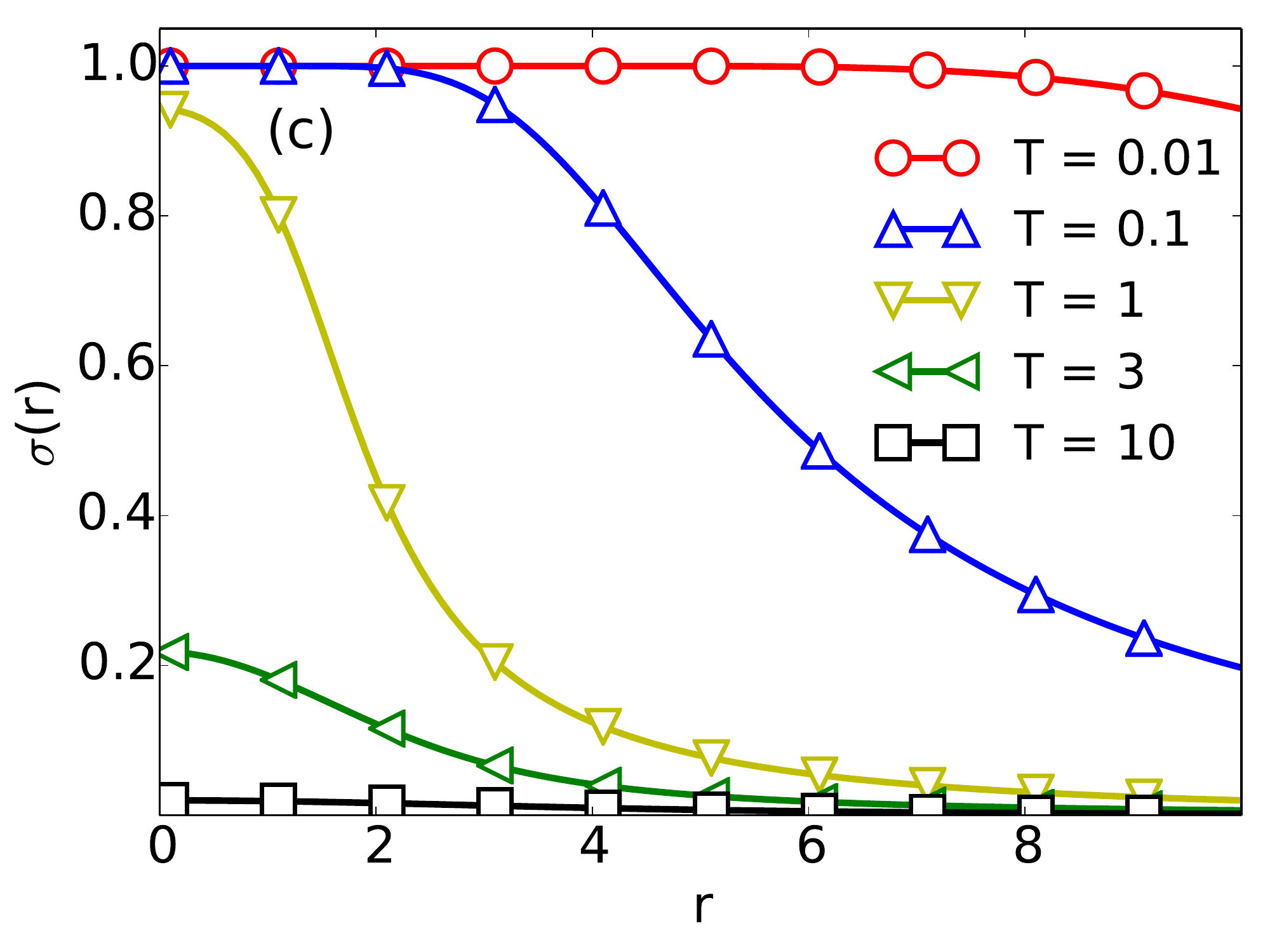}%
              \label{Fig2.4c}%
           }
           \caption{(Color online) Mean field predictions of (a) the number of H-bonds, $n_H(r)$, as a function of the distances from the ion at different temperatures.
           (b) Number of H-bonds in bulk at different temperatures. (c) The dipole moment at a site as a function of the distances from the ion at different temperatures.}
           \label{Fig2.4}
 \end{figure*}

We can obtain a mean field (MF) estimate of the number of H-bonds at a given site from formula \eqref{eq2.1}. Mean field involves replacing a
quantity by its corresponding mean and neglecting correlations. In this way we get a contribution of $\frac{1}{2}\left(\langle\sigma_{i}\rangle\langle\sigma_{j}\rangle + 1\right)$ to $n_H$ for each of the 
nearest neighbors from the equation \eqref{eq2.1}. The MF solution of the average dipole moment at a distance $r$ from the ion is given by the well known mean field 
solution from Ising model $\langle\sigma(r)\rangle = \tanh(\langle\sigma(r)\rangle+g(r))$, where
$g(r)\sim -\frac{\Gamma}{r^2}(1-e^{-x^2/\lambda^2})$ is the ion-dipole interaction at a temperature $T$. To first order approximation in the MF limit the number of H-bonds becomes
$n_H(r) \approx \tanh\left(\tanh(g(r))+g(r)\right)$. Figure \ref{Fig2.4}-(a) shows the behavior of the H-bonds in MF as a function of distance at various temperatures.
At very low temperatures all the water dipoles are aligned and H-bonded. The maximum number of H-bonds at a given site can be 4. As we lower the temperature
the orientation correlations becomes lower and the number of H-bonds decreases. Figure \ref{Fig2.4}-(b) shows the number of H-bonds in bulk at each site and it is 
well known it goes down with temperature.  Figure \ref{Fig2.4}-(c)
captures the mean field limit of the orientational/dipole order at various temperatures.

\section{Hydration - cation vs anion}
\label{Sec3}
Unlike the classical uniform dielectric model, hydration effect depends on the sign of the ion because the water molecule is an asymmetric dipole. 
In equation~\eqref{eq1.3} the Hamiltonain can not differentiate between the positive and negative charges of the ion as it depends on the magnitude of $\Gamma$,
the strength of the ion not its sign. So we modify the ion-water interactions to consider the ion-Oxygen and ion-Hydrogen interactions separately. Thus the total Hamiltonian becomes
\begin{align}
&H_N  = J/2\sum_{i,j}\left(\sigma_{i+1,j}^x\sigma_{i, j}^x + \sigma_{i,j+1}^y\sigma_{i, j}^y\right) + \sum_{i, j}\frac{\Gamma/2}{\sqrt{(i+\sigma^x_{i,j}b\sgn(i-N/2) - N/2)^2 + 
(j-N/2)^2}} \nonumber \\& +  \sum_{i, j}\frac{\Gamma/2}{\sqrt{(i - N/2)^2 + (j+\sigma^y_{i,j}b\sgn(j-N/2)-N/2)^2}} - \sum_{i, j}\frac{\Gamma}{\sqrt{(i - N/2)^2 + (j-N/2)^2}}
 \nonumber \\& + N^2(\epsilon_H-R) 
\end{align}
with $J = \epsilon_H + R$ and $b$ is the distance between the Oxygen and Hydrogen atoms in each water molecule. 
\begin{figure}[h]
        \centering
           \subfloat{%
              \includegraphics[height=4.2cm]{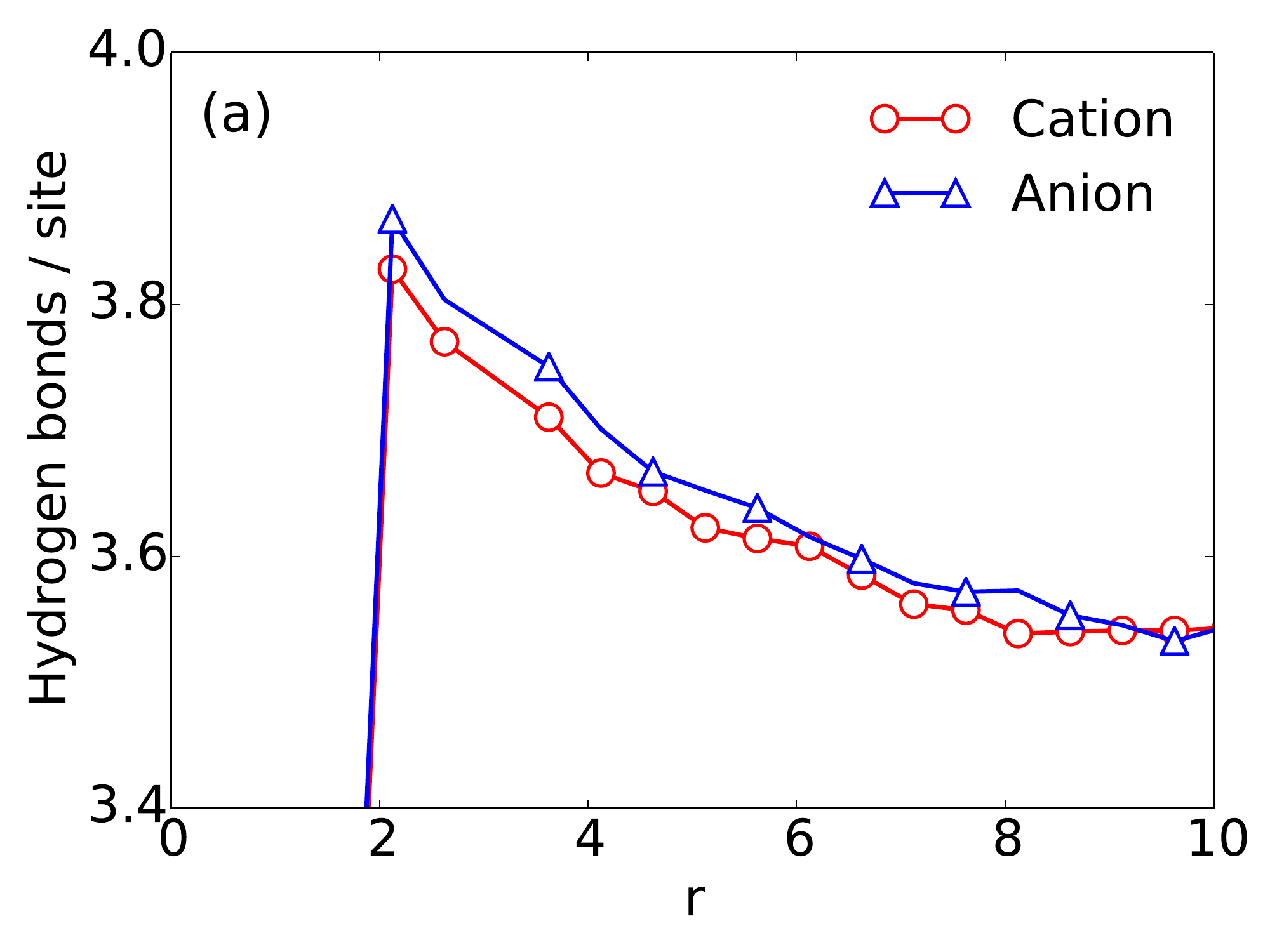}%
              \label{Fig3.1a}%
           }
           \subfloat{%
              \includegraphics[height=4.2cm]{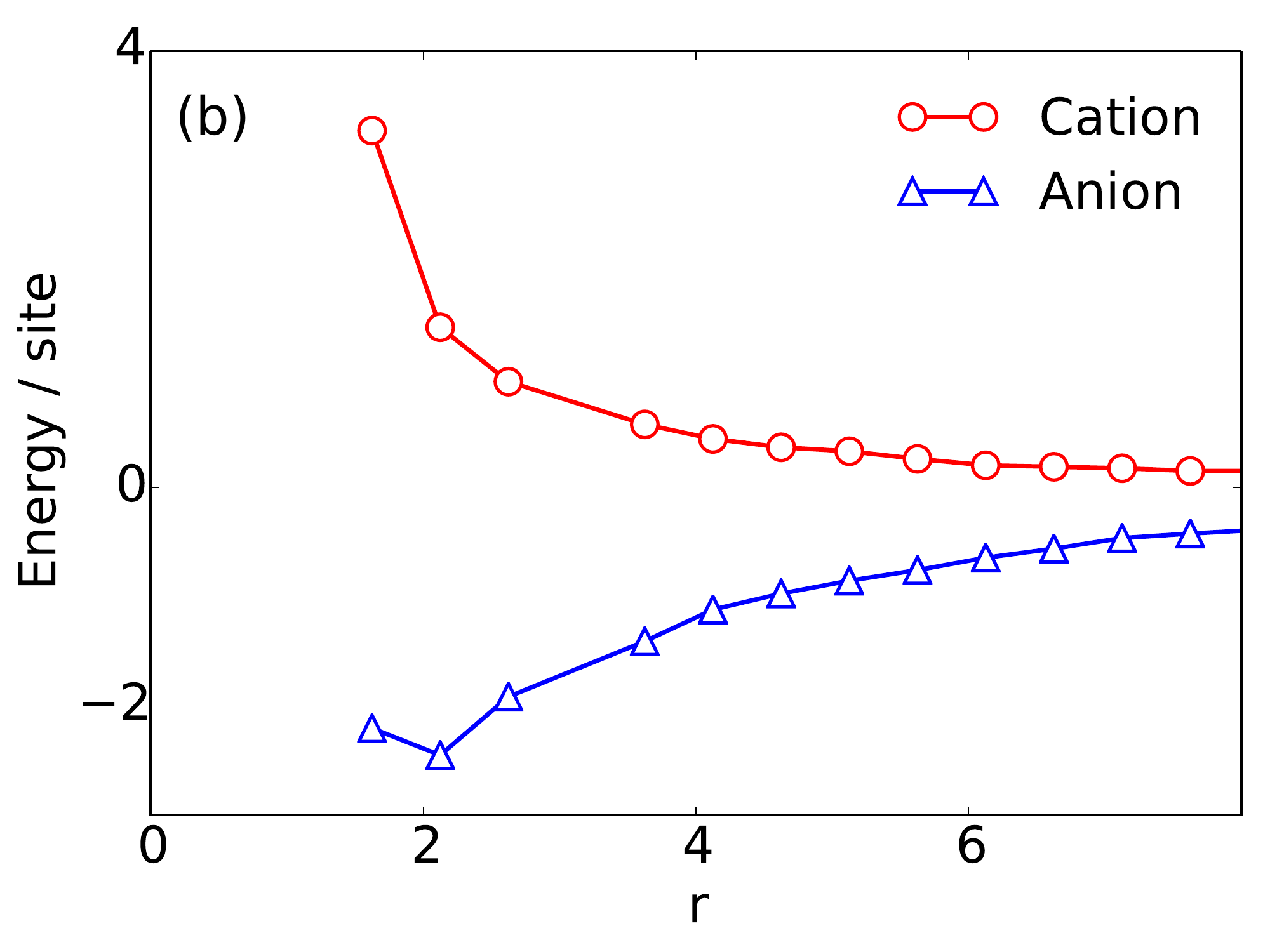}%
              \label{Fig3.1b}%
           }
            \subfloat{%
              \includegraphics[height=4.2cm]{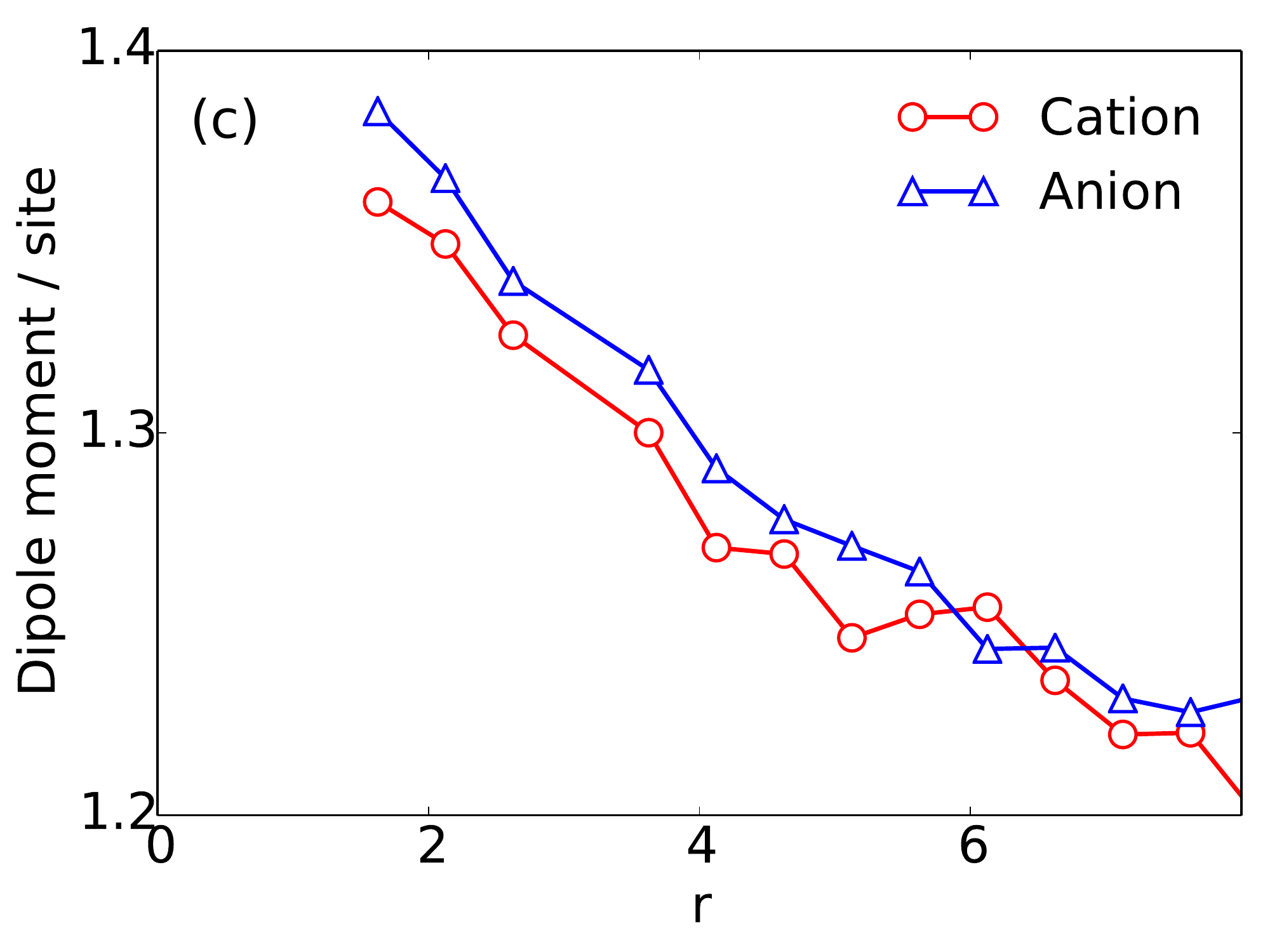}%
              \label{Fig3.1c}%
           }
           \caption{(Color online) The variation of (a) H-bonds, (b) insertion energy (c) dipole moment per site for anions and cations vs distance of the site
           from the ion at $T = 3$ and $\Gamma = 10$.}
           \label{Fig3.1}
 \end{figure}
We see from Figure \ref{Fig3.1}-(a) that anions have higher number of H-bonds and degree of orientation order than cations of same radius and charge.
Also the insertion energy is lower for anion than cation as a result anions are strongly hydrated compared to cations. The general behavior agrees with the reference 
\cite{Collins}. In the reference~\cite{Collins} this results from the quantum mechanical effect for which the effective interaction between 
anion and Hydrogen bond is stronger than cation and Oxygen. In our model, this is because of the asymmetry between positive and negative charges within the water molecule.
In this respect it is quite similar to the mechanism described in Ref \cite{lynden1997hydrophobic} where the outermost location of the positive charges in the 
water molecules allows them easier to closer to the anions than the negative Oxygen site to the cations.

\section{Hydration shells vs $\Gamma$ and radius of the ion}
\label{Sec4}

\begin{figure}[h]
        \centering
           \subfloat{%
              \includegraphics[height=6.2cm]{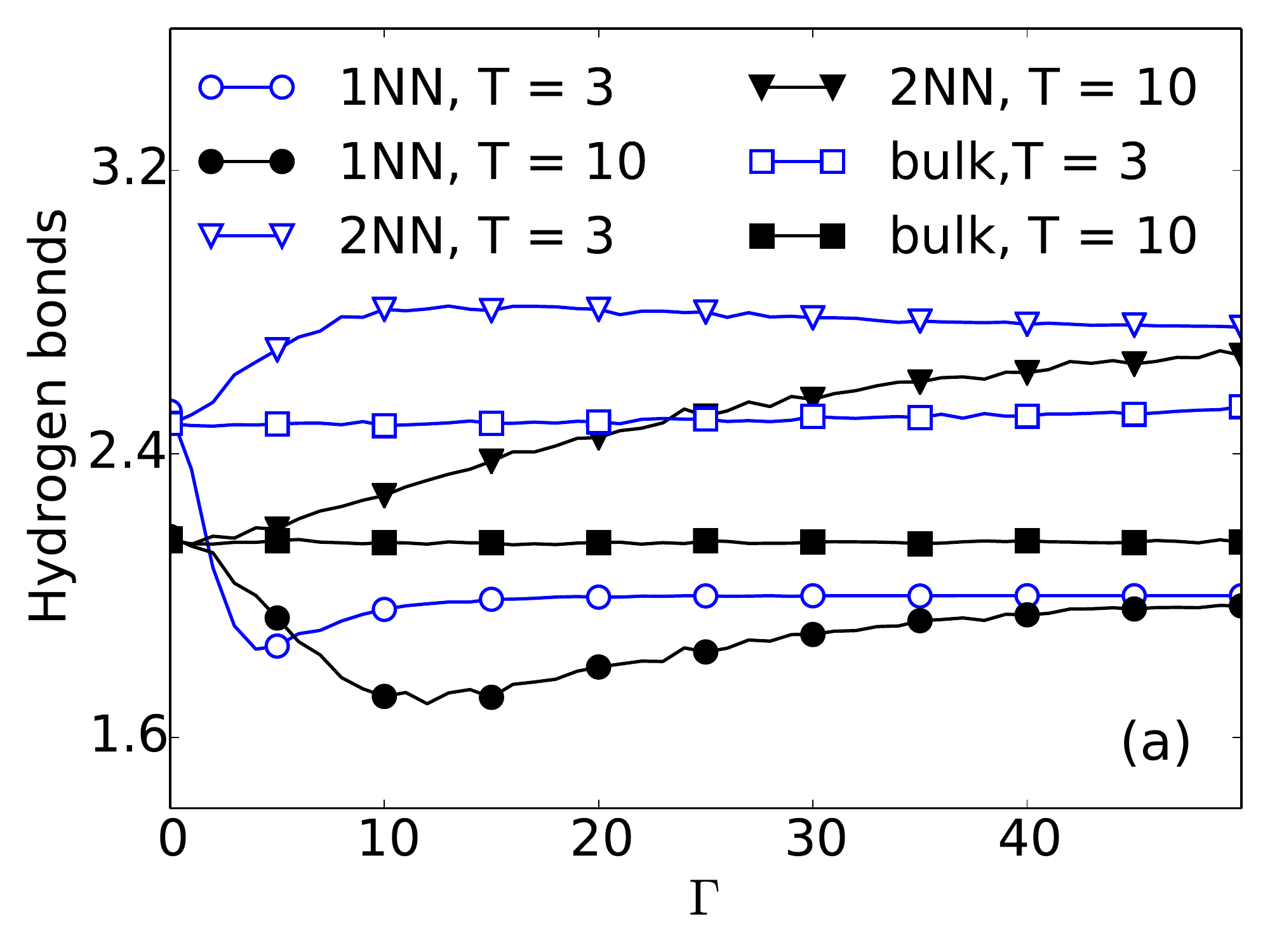}%
              \label{Fig4.1a}%
           }
           \subfloat{%
              \includegraphics[height=6.2cm]{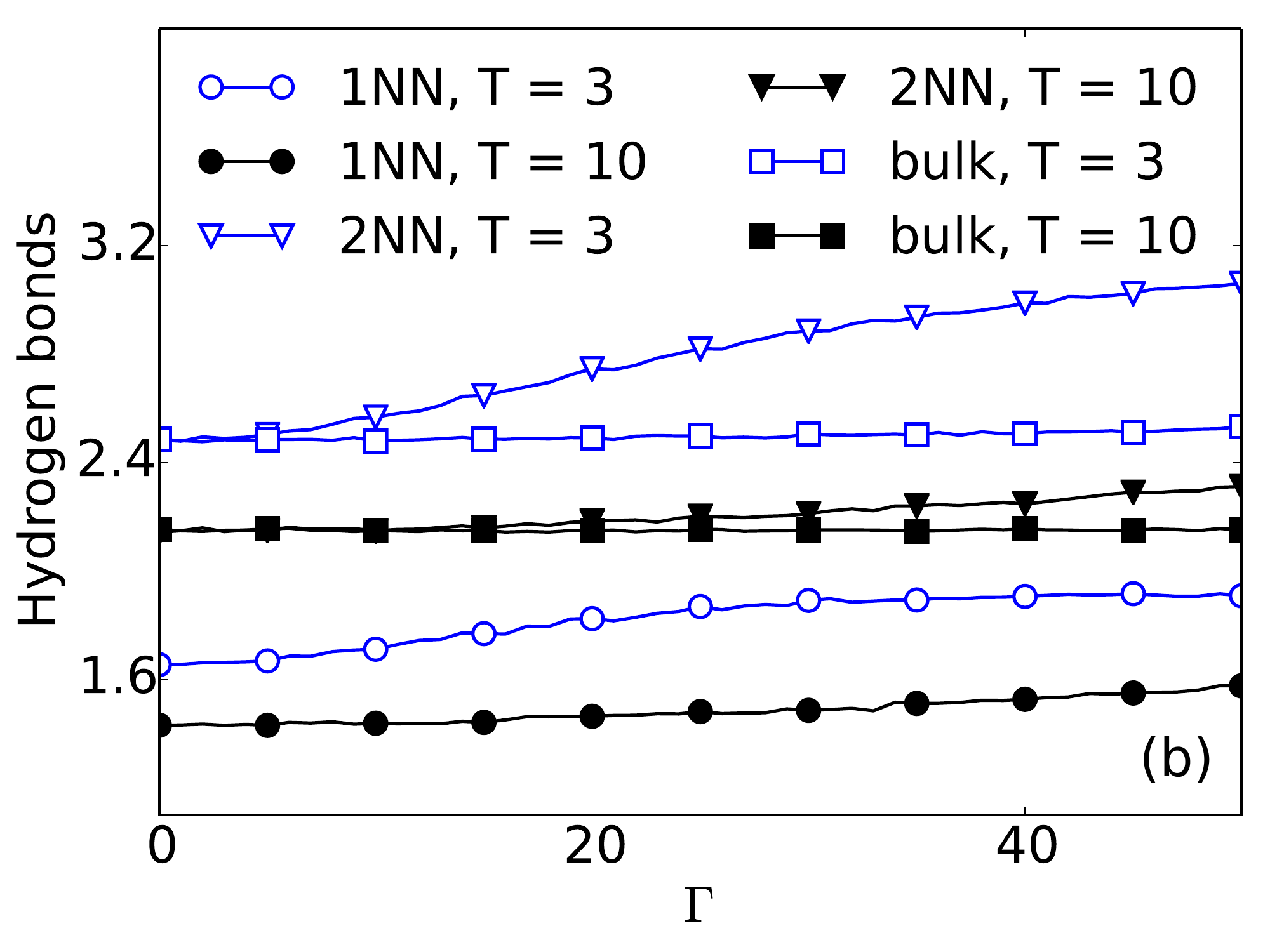}%
              \label{Fig4.1b}%
           }
           \caption{(Color online) The number of Hydrogen bonds vs $\Gamma$ for ion radius (a) $r_0 = 0$ and (b) $r_0 = 2$ at different temperatures.
           1NN denotes the water molecules closest to the ion, 2NN next nearest neighbor to the ion and the bulk.}
           \label{Fig4.1}
 \end{figure}

 \begin{figure*}[h]
        \centering
           \subfloat{%
              \includegraphics[height=6.2cm]{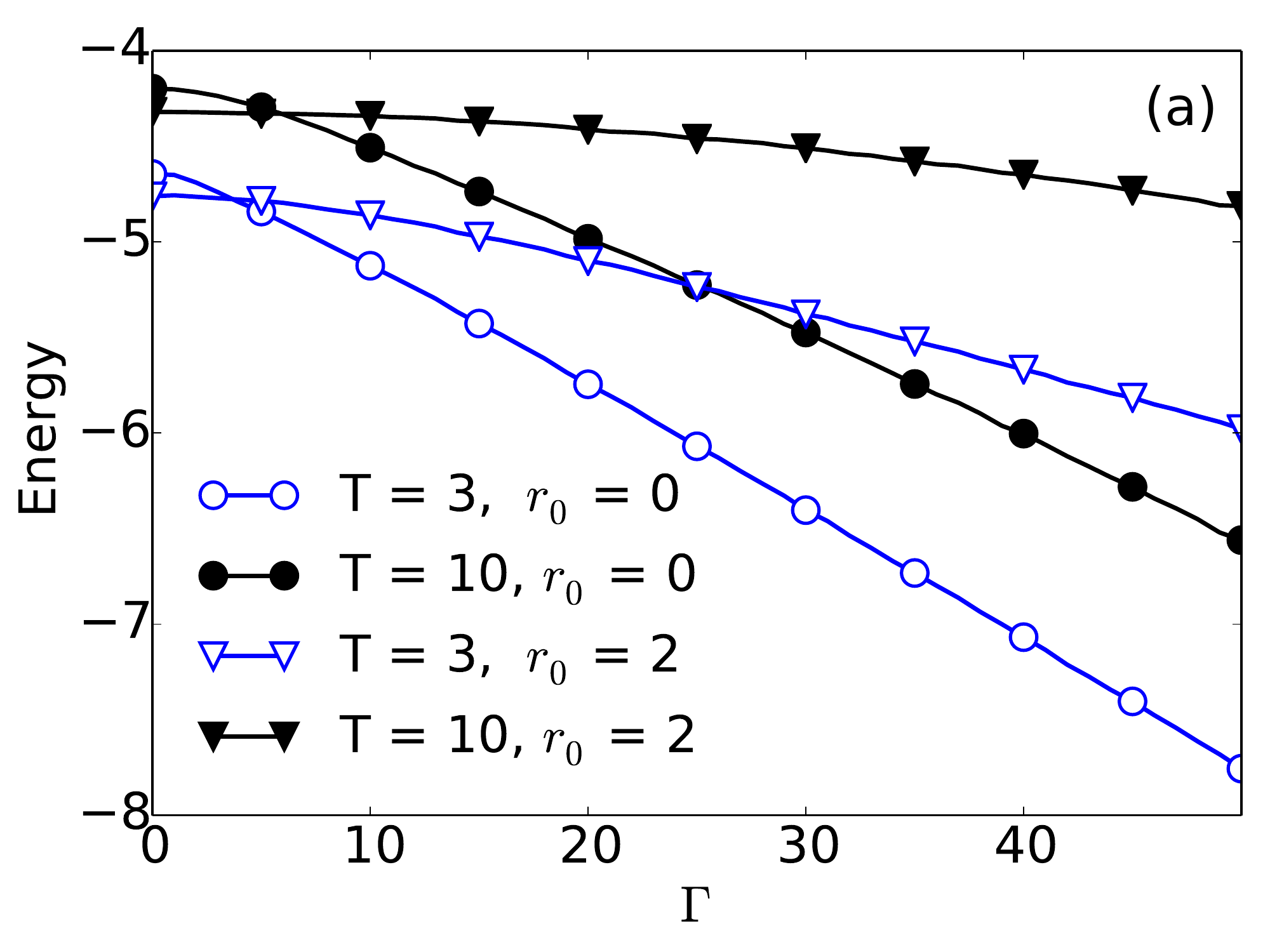}%
              \label{Fig4.2a}%
           }
           \subfloat{%
              \includegraphics[height=6.05cm]{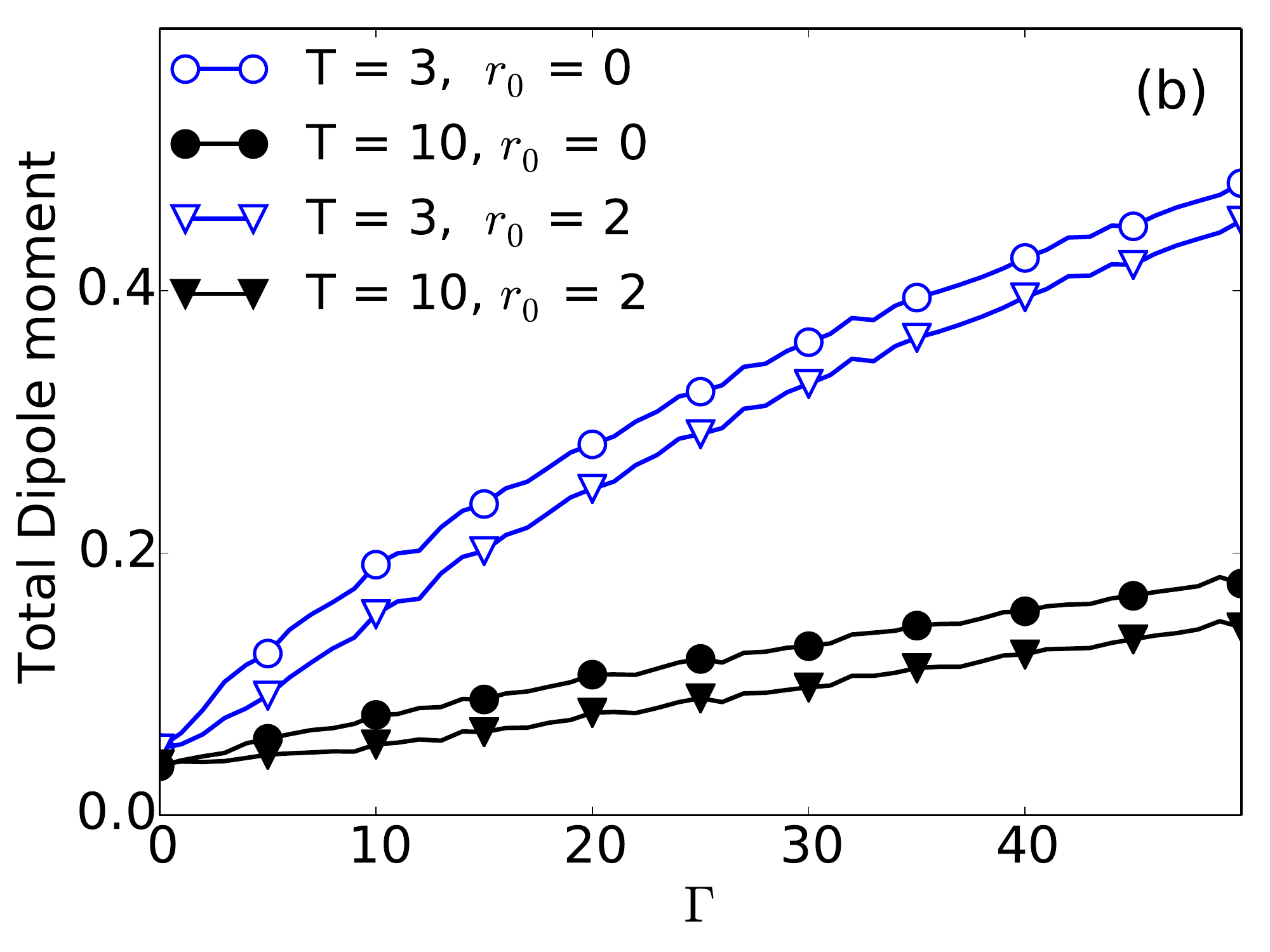}%
              \label{Fig4.2b}%
           }
           \caption{(Color online) (a) The insertion energy and (b) the dipole moment of water molecules vs $\Gamma$ for different values of the temperature and the ion radius.}
           \label{Fig4.2}
 \end{figure*}
At a given temperature increasing the charge on the ion has different effects on the H-bond network for smaller or larger ions. Figure \ref{Fig4.1} shows
that the H-bonds for the first nearest neighbors ( 1NN ), second neighbors ( 2NN ) and bulk are affected differently as we change the charge of the ion $\Gamma$.
For the ion of radius zero (point charge) in Figure \ref{Fig4.1}-(a), the decrease in 1NN H-bonds at low $\Gamma$ values corresponds to the reorientation of the water molecules resulting in the breaking of the 
pre-existing H-bonds. As we increase $\Gamma$ further they start forming H-bonds in the direction of the ions. For the next nearest neighbors, the first layer acts like a 
biased boundary condition, and it is easier to establish the H-bond with the first layer.
As a result, the number of H-bond at this layer is larger than that in bulk. The bulk layers are not affected by the ion. 
In Figure~\ref{Fig4.1}-(b) shows that the influence of ions on the first water layer is different from the second layer. 
There is a strong influence on the H-bond network in first layer regardless of the strength of the Coulomb interaction $\Gamma$. This decreases slightly on increasing $\Gamma$
because the strong ordering due to the ion can lead to strong H-bonding between the first layer and the second layer. This can be seen as an increase in the H-bonds in 2NN. 
When $\Gamma$ is small, which corresponds to small valency or large effective size of the ion, there is no influence to the 2NN. This is consistent with the experiments~\cite{marcus2009effect}.
Larger $\Gamma$, which corresponds to the larger valency of the ion or polarizability, induces stronger H-bond network in 2NN. This is highly correlated with the increase in
the number of H-bonds in 1NN as we noted. Our simulation results in Figure \ref{Fig2.2}-(c) agree with our findings in Figure \ref{Fig4.1}-(b). For the first shell the number of
H-bonds for $Ca^{2+}$ is almost the same with $Na^{+}$ similar to the slight increase in number of 1NN H-bonds with $\Gamma$ in Figure \ref{Fig4.1}-(b). The rate of increase
in H-bonding vs $\Gamma$ is sharper for 2NN in Figure \ref{Fig4.1}-(b), the trend that is repeated in Figure \ref{Fig2.2}-(c) where the number of H-bonds is appreciably higher in the second shell in $Ca^{2+}$ than in $Na^{+}$.

The insertion energy becomes more and more negative as we increase the $\Gamma$ as shown in Figure \ref{Fig4.2}-(a), thus it is easier to hydrate highly charged ions
compared to smaller $\Gamma$ ones. The Figure also shows that it harder to hydrate larger ions. Smaller ion size and larger ion charge
increases the orientational order of the water molecules as shown in Figure \ref{Fig4.2}-(b).

 \begin{figure*}[h]
        \centering
           \subfloat{%
              \includegraphics[height=6.2cm]{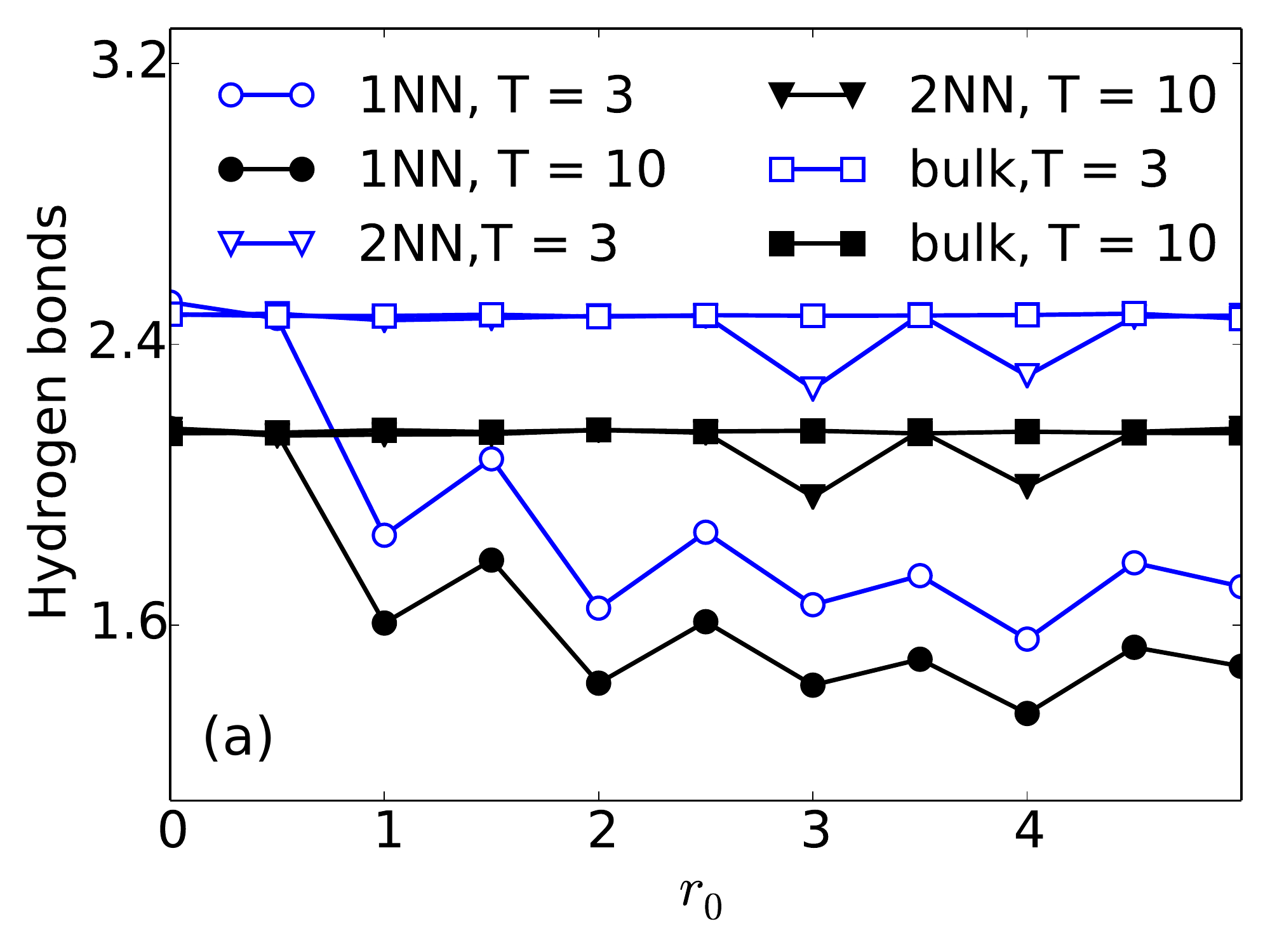}%
              \label{Fig4.3a}%
           }
           \subfloat{%
              \includegraphics[height=6.2cm]{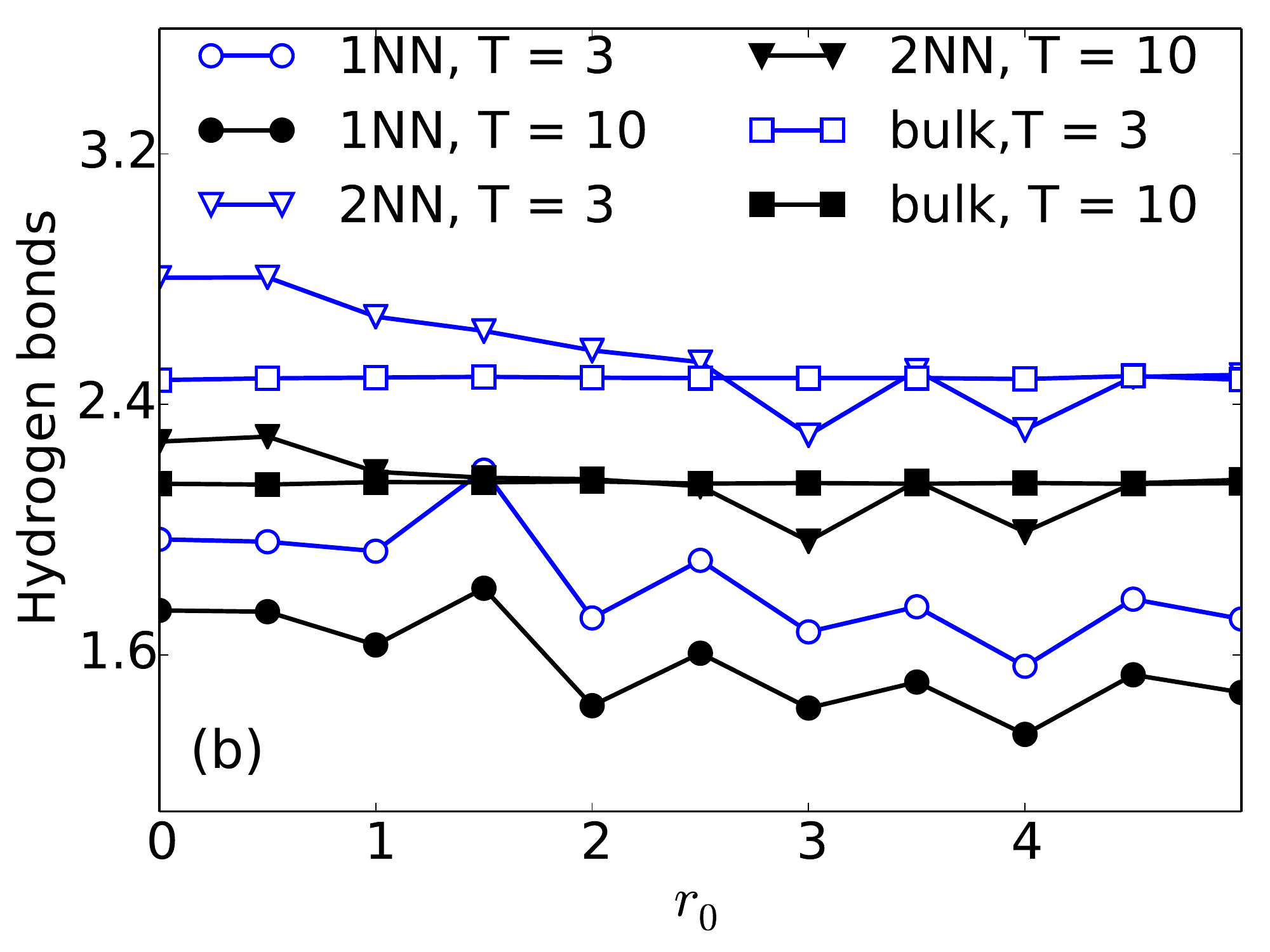}%
              \label{Fig4.3b}%
           }
           \caption{(Color online) The number of Hydrogen bonds vs ion radius $r_0$ for strength of the ion-dipole interactions (a) $\Gamma = 0$ and 
           (b) $\Gamma = 10$ at different temperatures. 1NN denotes the water molecules closest to the ion, 2NN second nearest neighbor 
           and the bulk water molecules.}
           \label{Fig4.3}
 \end{figure*}

 \begin{figure*}[h]
        \centering
           \subfloat{%
              \includegraphics[height=6.2cm]{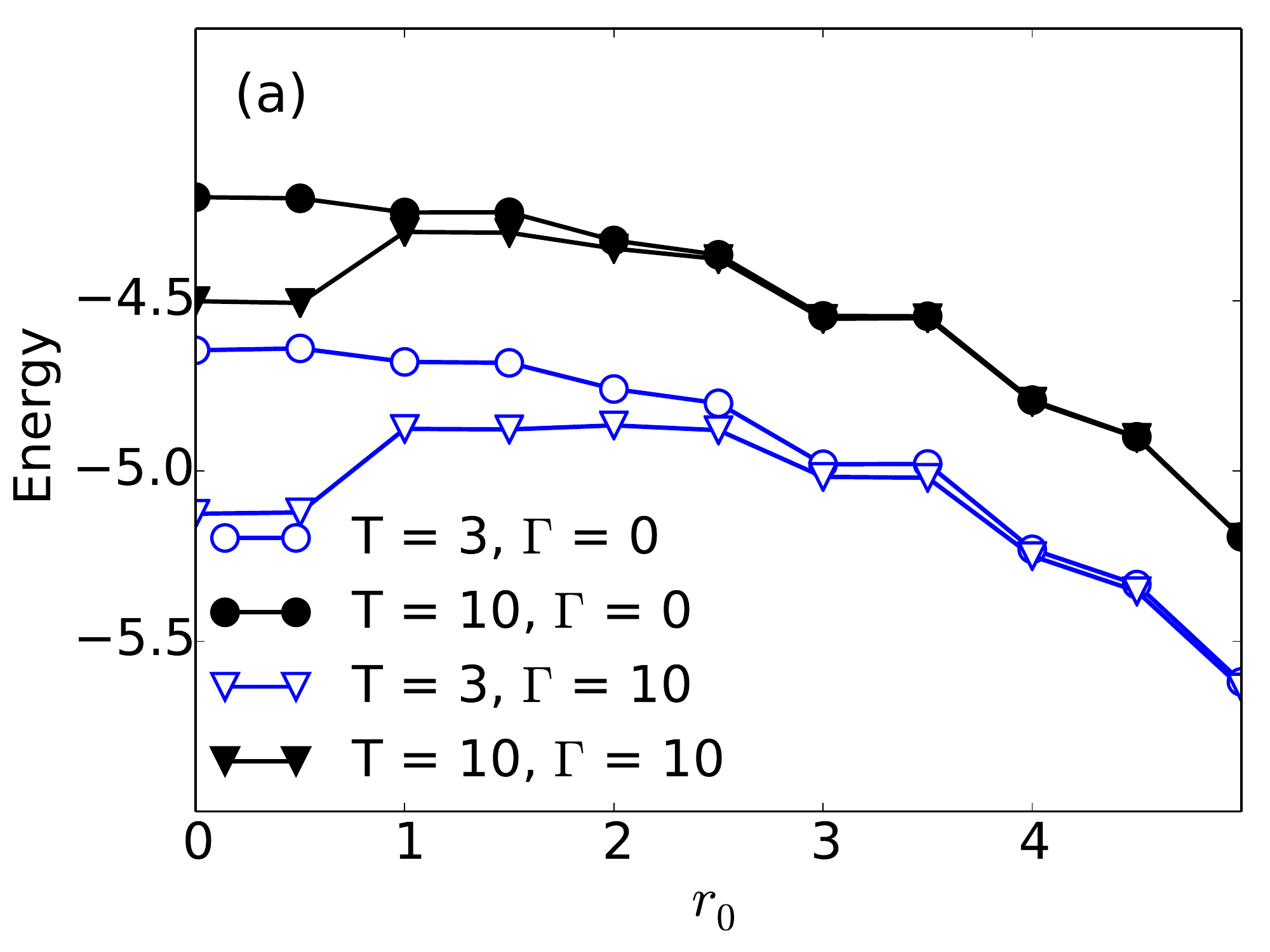}%
              \label{Fig4.4a}%
           }
           \subfloat{%
              \includegraphics[height=6.2cm]{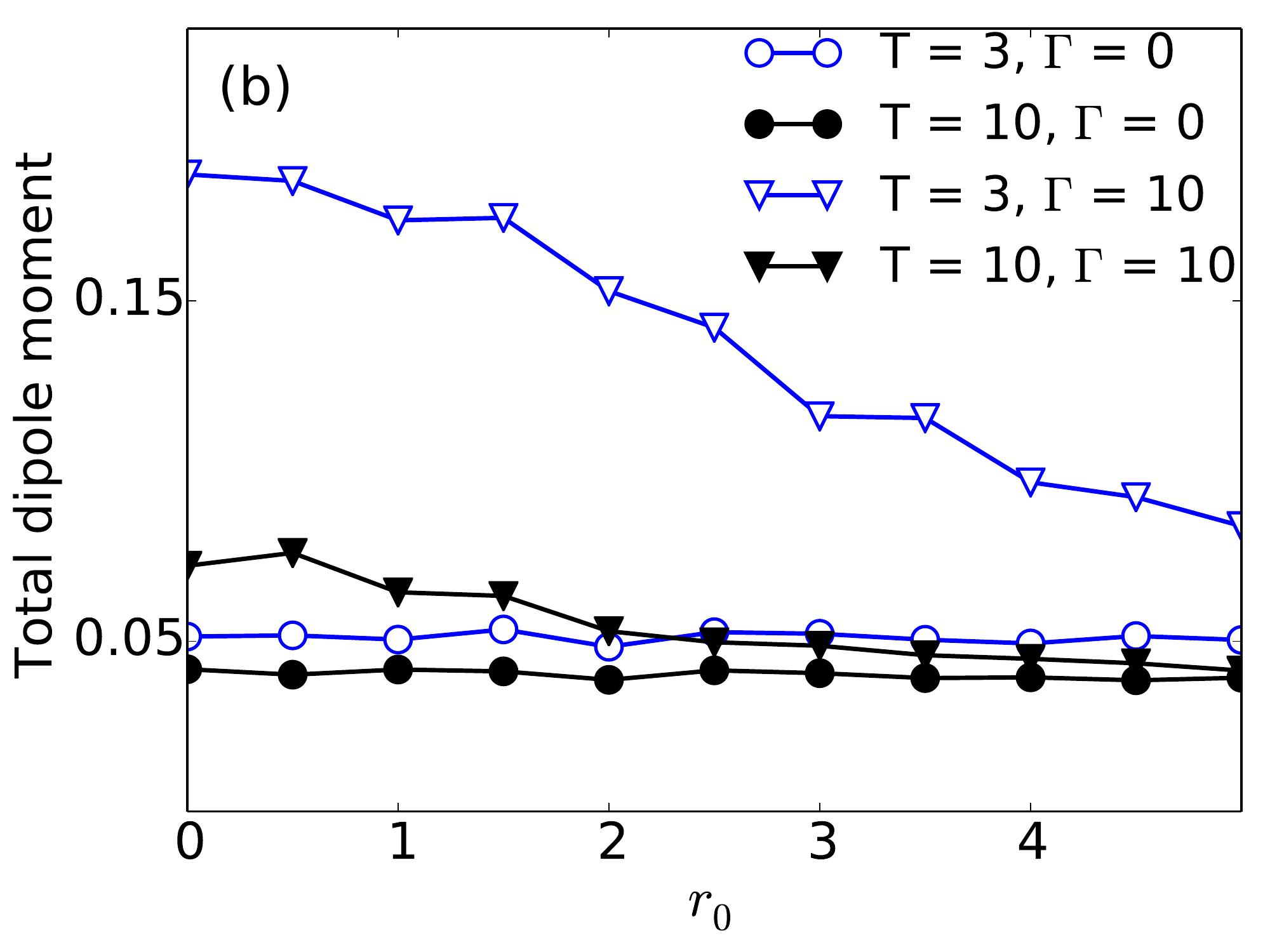}%
              \label{Fig4.4b}%
           }
           \caption{(Color online) (a) The insertion energy and (b) the dipole moment of water molecules vs ion radius for different values of temperature $T$ and $\Gamma$.}
           \label{Fig4.4}
 \end{figure*}
 
Ions affect the water properties in two different ways, as a defect to create cavity and as a source of electrostatic potential. 
Figure \ref{Fig4.3} shows the two effects separately. Figure \ref{Fig4.3}-(a) is for the case when the ion has no electrostatic interaction \textit{i.e.} 
it acts as a defect. Very small defect doesn't make any difference to the H-bond structure. Larger defects decrease the number of H-bond in 1NN because 
of the reduction in H-bonding sites due to the defect. Interestingly, the number of H-bonds for 2NN has a very weak dependence on the radius of the ion 
once the electrostatic interactions are switched off. This means that the overshooting of H-bonds in 2NN in Figure \ref{Fig2.2} compared to the bulk is 
consequence of the electrostatic interactions via modifying the configuration of the water at 1NN. 
Comparison with Figure \ref{Fig4.3}-(b) shows that electrostatic interactions increase the number of H-bonds in 2NN. On increasing the radius of the ion, the number 
of H-bond converges to bulk value which confirms the idea that the larger H-bond in 2NN is a combined result of the biased network in 1NN from the
electrostatic interactions. Thus for large ions the role of the ion as a defect has a stronger influence on the water properties than the electrostatic interactions.
Since in our definition the number of H-bonds is measured by the orientational correlations, as the ion radius increases the number of effective neighbors 
for the first two water layers decreases and hence $n_H$ decreases. Simulation results for $K^+$ and $Na^+$ in Figure \ref{Fig2.2}-(c) support this. 

The insertion energy for $\Gamma = 0$ is the energy required to create a defect in the water. Figure \ref{Fig4.4}-(a) shows that it is energetically unfavorable 
to put an ion of large radius in water. For large ions the charge densities is small enough for the electrostatic interactions to play a role in their hydration.
The orientational order decreases rapidly with the radius of the ion size as shown in Figure \ref{Fig4.4}-(b).

\section{polarizability}
\label{Sec5}
\begin{figure*}
        \centering
           \subfloat{%
              \includegraphics[height=6.2cm]{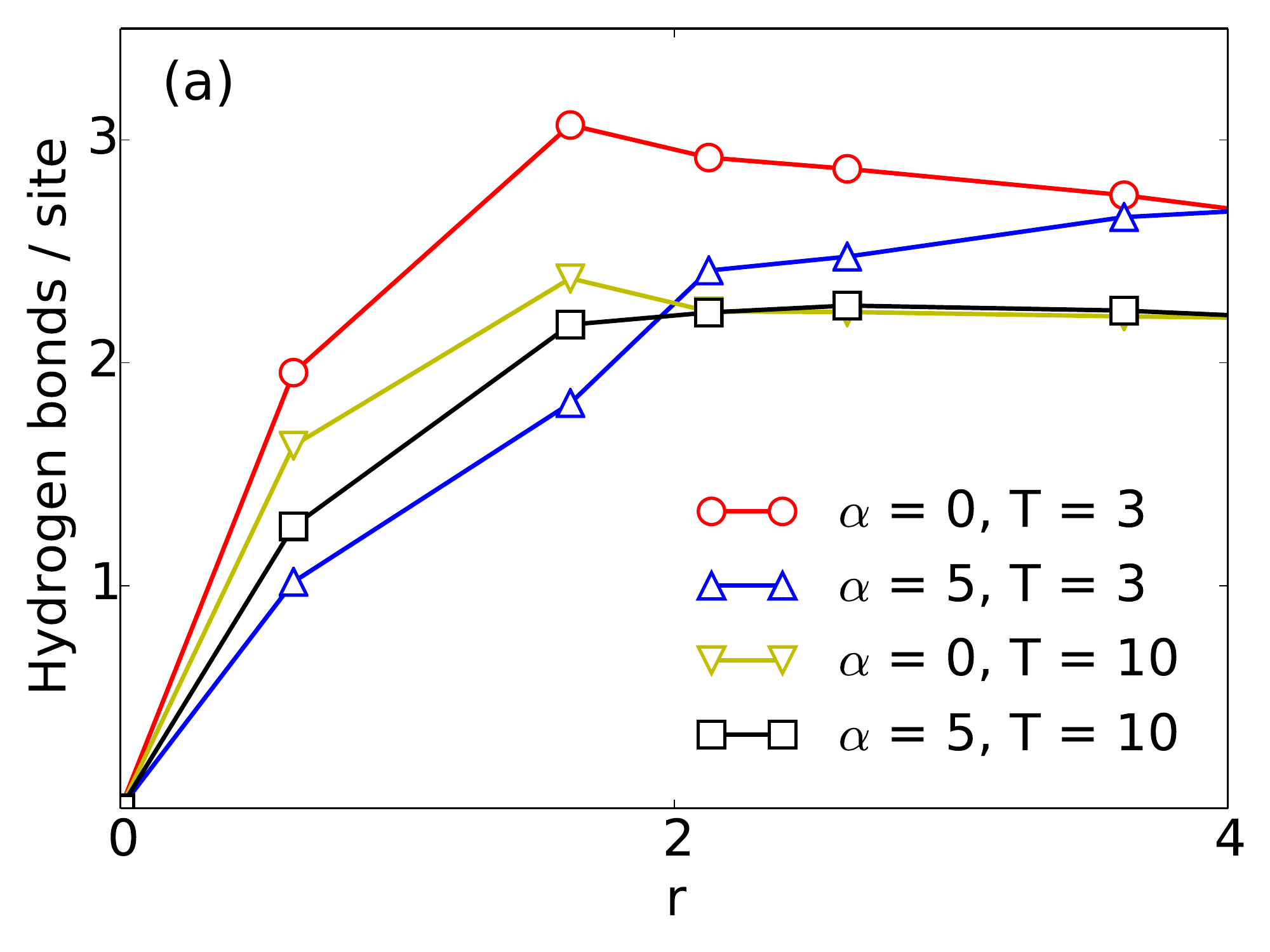}%
              \label{Fig5a}%
           }
           \subfloat{%
              \includegraphics[height=6.2cm]{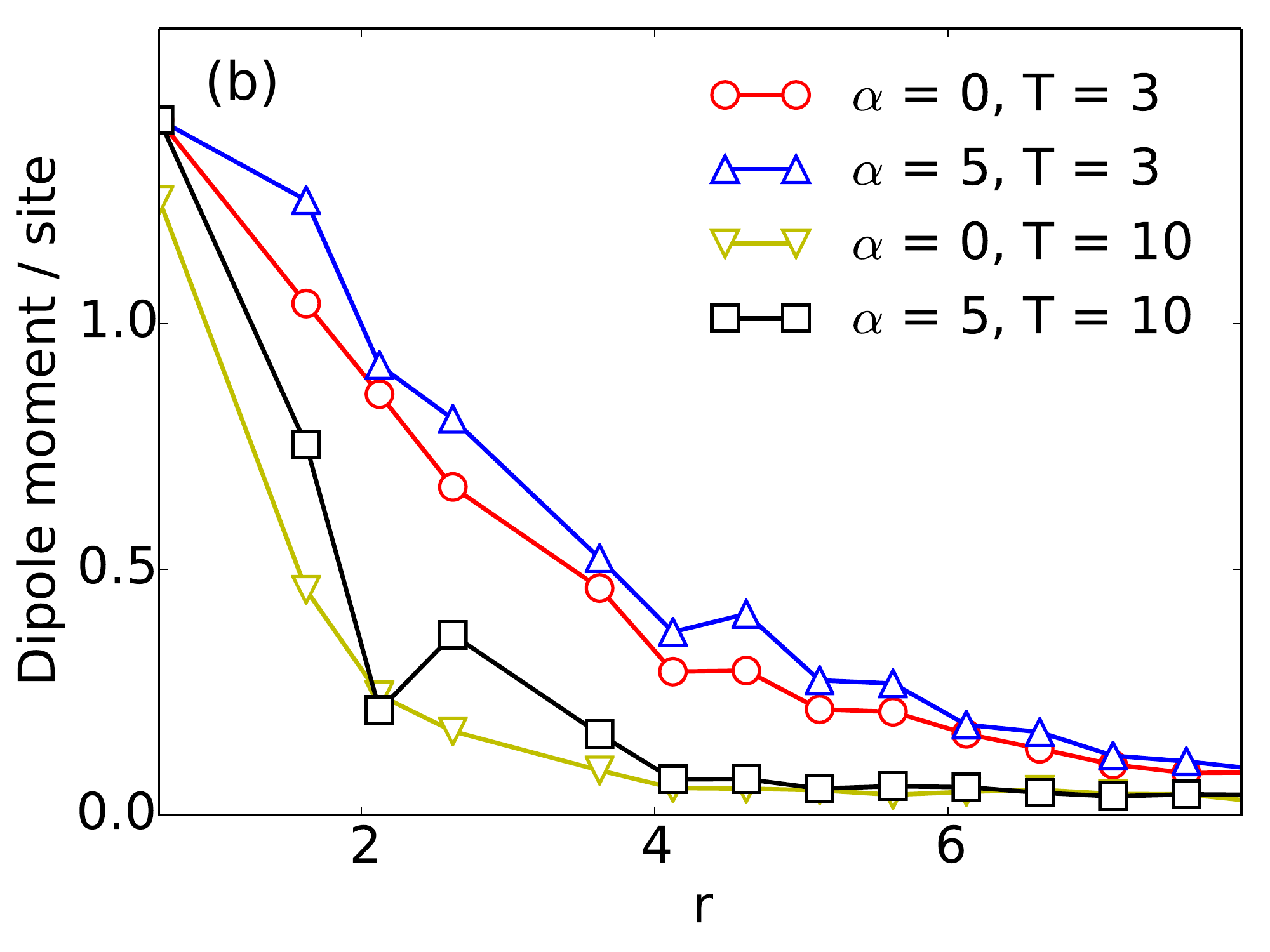}%
              \label{Fig5b}%
           }
           \caption{(Color online) (a) The total number of hydrogen bonds and (b) the total dipole moment of the water molecules vs the ion-water site distance $r$ 
           for $\Gamma = 10$ at different temperatures and polarizations $\alpha$.}
           \label{Fig5}
 \end{figure*}
 
One interesting property of water is the polarizability of the water molecules which distorts the charge distribution of the water molecules in presence of ions. As reported 
in some references the polarization effect is crucial to specific ion effect~\cite{dos2010surface,ding2014anomalous}. Depending on the consideration of polarizability
of the ions themselves the trend of the specific ion effect can be opposite. Here we investigate how the polarizability of water molecules influences our model. 
The polarization induces a dipole moment in the water molecules proportional to the electric field of the ion in addition to the permanent dipole moment of the water molecules. 
To include the effects of polarizability, we modify the strength of the interactions between the water molecules at site $(i ,j)$ in the equation \eqref{eq1.3} by 
$J_x(i, j) = J + \alpha_xg_x(i, j)$ and $J_y(i, j) = J + \alpha_yg_y(i, j)$. $\alpha_x$ and $\alpha_y$ are the polarization along the $x$ and $y$ directions 
respectively. 
From the Figure \ref{Fig5}-(a) we see that the hydrogen bonds per site decrease on increasing the polarization of the water molecules. 
It indicates that the polarization induced dipolar interactions are dominant over the energy of H-bond network at short distances, 
especially at low temperatures. At high temperatures the hydrogen bonds per site converges to the non-polarizable model at large distance quickly. 
In Figure \ref{Fig5}-(b) we plot the total dipole moment of the water molecules at different temperatures and polarizabilities.
Increasing the polarizability increases the total dipole moment compared to the unpolarized case which corresponds to the bare permanent dipole moment case.  

\section{Conclusions and Discussions}
We present a simple lattice model of water based on 2D Ising model to capture the competing interactions of the short-ranged H-bonds and the long-ranged 
electrostatic interactions in the presence of an ion. This simple model can explain many conjectures, and experimental results about the hydration effect, 
especially on the stability of H-bonds near the ions~\cite{Collins,marcus2009effect}.
Our formalism explains qualitatively the dependence of the hydration properties of ion charge and radius, the role of the ion in breaking of the H-bond network and 
finally the differences between the hydration of cations and anions. 
We do not consider the change in coordination number depending on distance between the ion and the water molecules or the translational entropy of the water molecules. 
The angular distribution is also discretized, and because of this the competition between H-bond and dipolar order is not well included. This can be improved in a 3D 
lattice model. We also omitted quantum effects which could be important to the many body interactions in water. 
In spite of these weaknesses, our model gives intuition for the very complicated effects of ion hydration. Its strength lies in its simplicity and extensibility.
We can extend it to many body system, or three dimension in straight forward way, although obtaining an analytic solution would not be easy. 
We also can consider the proton hopping and the ionization or deionization of water molecules which is very important in many systems. It can be also applied to the 
study of the overlapping of hydration layer and their co-operativity. These would be done in a subsequent paper. 

\section{Acknowledgement}

This work was supported by the Ministry of Education, Science, and Technology (NRF-2012R1A1A2009275, NRF-C1ABA001-2011-0029960) of the National Research Foundation 
of Korea (NRF), and the National Institute of Supercomputing and Network/Korea Institute of Science and Technology Information with supercomputing resources including 
technical support (KSC-2015-C1-009), and PLSI supercomputing resources of Korea Institute of Science and Technology Information.

\bibliography{water}
 
%

\end{document}